\documentclass[12pt]{iopart}
\usepackage{enumerate}
\usepackage{bbm}

\expandafter\let\csname equation*\endcsname\relax
\expandafter\let\csname endequation*\endcsname\relax
\usepackage{amsmath}

\pdfoutput=1
\usepackage{braket}
\usepackage[paperwidth=210mm,paperheight=297mm,centering,hmargin=2cm,vmargin=2.5cm]{geometry}
\usepackage{relsize}
\usepackage[pdftex]{graphicx}
\usepackage[colorlinks=true,citecolor=blue]{hyperref}
\usepackage{cite}
\usepackage{mathrsfs}
\usepackage{array}
\usepackage{caption}
\usepackage{ulem}
\usepackage[symbol]{footmisc}

\usepackage[colorinlistoftodos]{todonotes}
\usepackage{multirow}
\usepackage{bbold}
\usepackage{braket}
\usepackage{float}
\usepackage[caption = false]{subfig}

\usepackage[english]{babel}
\usepackage[utf8x]{inputenc}
\usepackage[T1]{fontenc}

\hypersetup{
   colorlinks=true,         % false: boxed links; true: colored links, false is default
   linkcolor=blue,          % color of internal links, red is default
   citecolor=blue,          % color of links to bibliography
}

\def\<{\langle}

\def\P{ {\cal P} }

\def\P{ {\cal P} }

\def\>{\rangle}
\def\<{\langle}

\newcommand{\be}{\begin{equation}}
\newcommand{\ee}{\end{equation}}

\newcommand{\zo}[1]{{\color[RGB]{0,0,0}{#1}}}

\begin{document}

\title{Gibbs mixing of partially distinguishable photons with a polarising beamsplitter membrane}

\author{Zo\"e Holmes}
\address{Controlled Quantum Dynamics Theory Group, Imperial College London, London, SW7 2BW, United Kingdom. \\ 
CEMPS, Physics and Astronomy, University of Exeter, Exeter, EX4 4QL, United Kingdom. \\ 
Information Sciences, Los Alamos National Laboratory, Los Alamos, NM, USA.}
\eads{zholmes@lanl.gov}

\author{Florian Mintert}
\address{Controlled Quantum Dynamics Theory Group, Imperial College London, London, SW7 2BW, United Kingdom.}

\author{Janet Anders}
\address{CEMPS, Physics and Astronomy, University of Exeter, Exeter, EX4 4QL, United Kingdom. \\
Institut f\"ur Physik und Astronomie, Potsdam University, 14476 Potsdam,  Germany.}

\date{\today}

\begin{abstract}

For a thought experiment concerning the mixing of two classical gases, Gibbs concluded that the work that can be extracted from mixing is determined by whether or not the gases can be distinguished by a semi-permeable membrane; that is, the mixing work is a discontinuous function of how similar the gases are. Here we describe an optomechanical setup that generalises Gibbs' thought experiment to partially distinguishable quantum gases. Specifically, we model the interaction between a polarisation dependent beamsplitter, that plays the role of a semi-permeable membrane, and two photon gases of non-orthogonal polarisation. We find that the work arising from the mixing of the gases is related to the potential energy associated with the displacement of the microscopic membrane, and we derive a general quantum mixing work expression, valid for any two photon gases with the same number distribution. The quantum mixing work is found to change continuously with the distinguishability of the two polarised gases. In addition, fluctuations of the work on the microscopic membrane become important, which we calculate for Fock and thermal states of the photon gases. Our findings generalise Gibbs' mixing to the quantum regime and open the door for new quantum thermodynamic (thought) experiments with quantum gases with non-orthogonal polarisations and microscopic pistons that can distinguish orthogonal polarisations.

\end{abstract}

\maketitle

\section{Introduction}

The importance of the concept of distinguishability in classical and quantum thermodynamics is neatly illustrated by Gibbs' mixing thought experiment~\cite{Gibbs}. In classical physics the work that can be extracted from mixing two gases is a discontinuous function of their similarity, determined by whether or not the gases can be distinguished by a semi-permeable membrane. In the quantum regime, the possibility of non-orthogonal quantum states allows one to consider gases that are neither perfectly distinguishable nor perfectly indistinguishable but rather at best partially distinguishable~\cite{Schrodinger, thesoviets, Lande1, Lande2, allahverdyan,Peres, Maruyama}. Thus the question arises, \textit{how does the work output from mixing quantum gases depend on their distinguishability?}

This question has been investigated from an information theoretic perspective. Early studies based on entropy arguments~\cite{Schrodinger, thesoviets, Lande1, Lande2} showed that 
the work output in the thermodynamic limit of large gases should increase continuously with the distinguishability of the two gases, quantified by the overlap in the gas particles' internal states. More recently~\cite{allahverdyan}, the Gibbs mixing of finite sized quantum gases has been shown to lead to an extractable work, defined as the mixing `ergotropy'~\cite{ergotropy}, that increases smoothly with distinguishability for homogeneous (made up of the same particles) gases, with more complex behaviour observed for inhomogeneous gases. 
However, the above approaches~\cite{Schrodinger, thesoviets, Lande1, Lande2, allahverdyan,Peres,Maruyama} do not discuss how the quantum gases and semi-permeable membranes (or alternative work extraction mechanisms) might be realised. Nor do they consider the time dependent dynamics of the mixing processes. Thus these information theoretic approaches leave open how such mixing processes could physically manifest.

In this paper we investigate a more concrete quantum generalisation of the Gibbs mixing thought experiment,
that may in principle be realised with an optomechanics setup~\cite{Elouard_2015,OptoMechExperiment,OptMechthermo1,OptoMechthermo2,OptoMechthermo3, OptoMechthermo4, OptoMechthermo5}.
Specifically, we will investigate the work that can be extracted from mixing two photon gases distinguished by their polarisation.
To study this we introduce a novel optomechanical setup that is similar to current experimentally realisable `membrane-in-the-middle' setups~\cite{experiment1, experiment1a,experiment2,experiment3,experiment4,experiment5, experiment6} and the thought experiment proposed in \cite{BunchingOptMech}, but uses a Polarisation dependent BeamSplitter (PBS) instead of the usual beamsplitter (BS). The PBS membrane we consider acts as a beamsplitter for one specific polarisation, say vertical, while as a mirror on the orthogonal polarisation, horizontal, thus realising a quantum version of a semi-permeable membrane.

In Section~\ref{sec:Gibbs} classical Gibbs mixing is reviewed before we introduce the PBS-in-the-middle optomechanical setup and its Hamiltonian in Section~\ref{sec:opto}. In Section~\ref{sec:Calc} we calculate the time dependent dynamics of the state of the photon gases and the PBS membrane and associate the motion of the membrane with a work output in Section~\ref{sec:OptMechMixResults}.
We find, as expected, that the mixing work increases smoothly with the distinguishability of the two photon gases, clearly extending the discontinuous  classical case and inline with previous results for quantum mixing~\cite{Schrodinger, thesoviets, Lande1, Lande2, allahverdyan,Peres,Maruyama}. 
However, with the explicit time dynamics for the optomechanical setting, we can characterise not just the average work but also build a more complete picture of the mixing process as discussed in Section~\ref{sec:fluctuations}.
For example, large fluctuations in the work output are found that arise from quantum fluctuations as well as classical fluctuations for initial Fock and thermal photon states. 
Experimental prospects of observing the Gibbs mixing work as a function of distinguishability are discussed in Section~\ref{sec:GibbsExp} before the findings are discussed in the conclusion section~\ref{sec:conc}.

\section{Classical Gibbs Mixing} \label{sec:Gibbs}

\begin{figure}[t]
\centering
\subfloat[Initial setup.]{\includegraphics[width=0.475\linewidth]{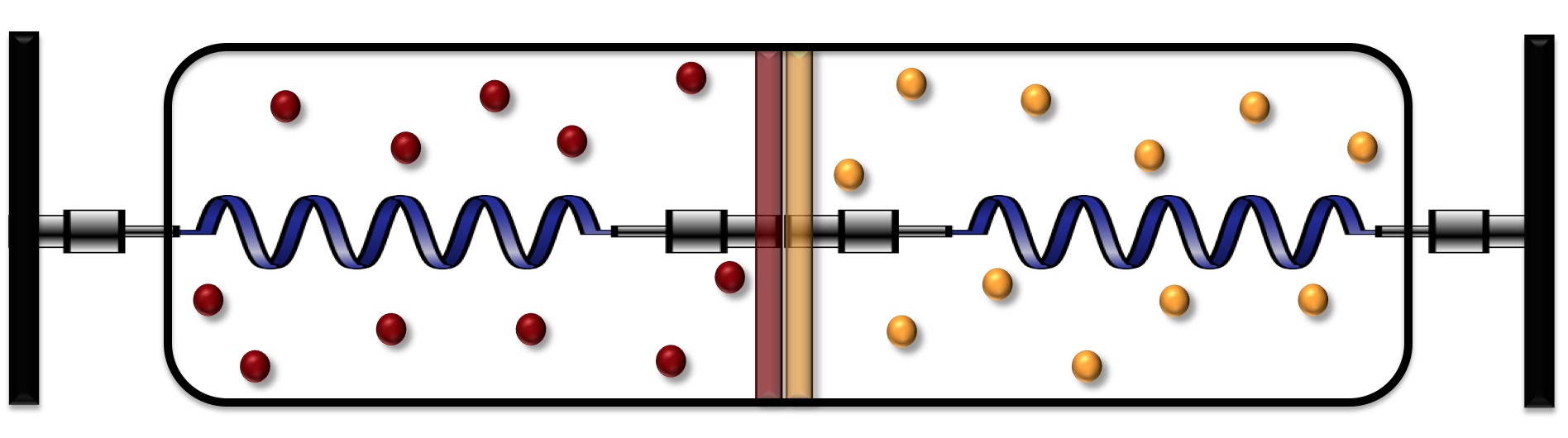}} \hspace{5mm}
\subfloat[Drawing work from mixing.]{\includegraphics[width=0.475\linewidth]{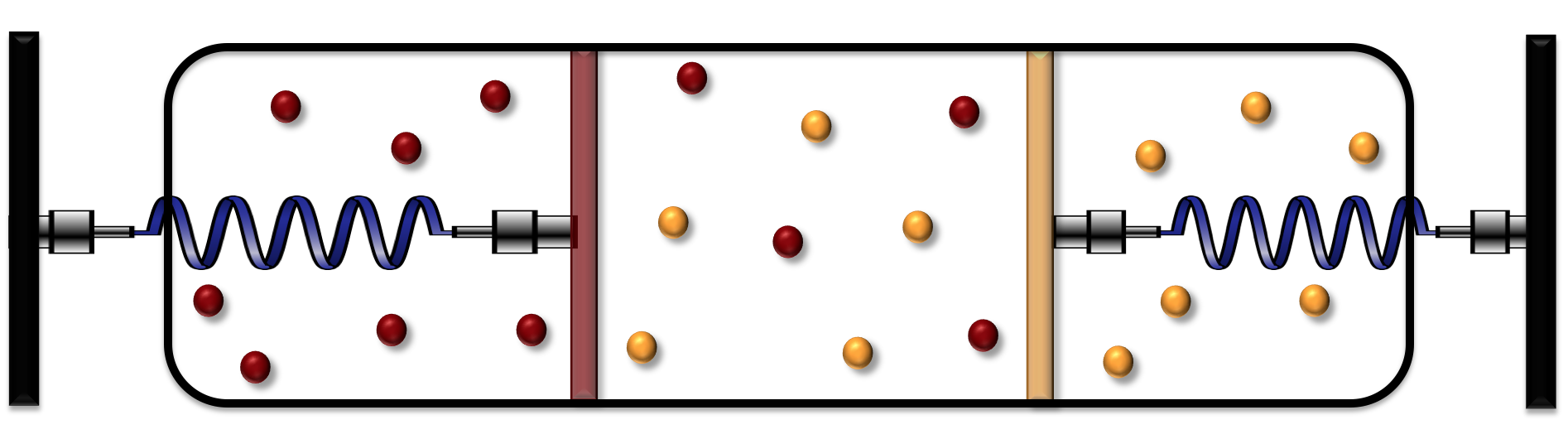}} 
\caption{\label{fig:classicalmixing} {\bf Gibbs mixing.} (a) Two homogeneous gases, a `red' gas and a `yellow' gas, are initially confined to two halves of a box of volume $2v$. To draw work from the mixing of the two gases, a pair of semi-permeable membranes is inserted, with red (yellow) gas particles passing through the red (yellow) membrane while being confined by the yellow (red) membrane. (b) The membranes are held in place with two springs, allowing them to move. Since each membrane confines one gas, which exerts a pressure on it, the membranes will slide and the gases compress the springs, performing work. The expansion of each of the two ideal gases from $v$ to $2v$, while in thermal contact with a heat bath at temperature $T$, thus mixes the gases and produces work while drawing heat from the bath.
}
\end{figure}

The standard classical protocol~\cite{Gibbs} for extracting work from mixing two homogeneous gases is sketched in Fig.~\ref{fig:classicalmixing}. 
Initially, there are $n$ particles of an ideal gas of type $L$ and $n$ particles of an ideal gas of type $R$, confined in two volumes, each of size $v$, separated by a pair of membranes. 
One of these membranes is permeable to type $L$ particles while impermeable to type $R$ particles, and vice versa for the second membrane. Both gases are in thermal equilibrium with an external heat reservoir at temperature $T$ and each gas exerts pressure, $p$, on its confining membrane, while exerting no pressure on the other membrane. By allowing the membranes to move under the force of the gases, the two gases can isothermally expand resulting in them mixing in the space between the membranes, see Fig.~\ref{fig:classicalmixing}. At the end of the protocol the two gases occupy the full volume $2v$ and are fully mixed. By standard thermodynamics, the work done on the membrane as the two distinguishable gases expand from volume $v$ to $2v$ is given by~\cite{Gibbs,Peres}
\begin{equation}\label{eq:distingwork}
    W_{\rm dist} = 2 \, \int_v^{2v} p \, dV = 2 n \, k_B T \ln 2 \ ,
\end{equation}
and the energy lost by the gases is  continually replenished as heat from the heat reservoir at temperature $T$. 
Crucially, this protocol is only possible if two suitable semi-permeable membranes can be found. Clearly, if the gases are in fact indistinguishable then no such membranes are available and thus no work can be drawn from  mixing of two indistinguishable gases, $ W_{\rm indist} = 0$. 

It has been argued by some~\cite{allahverdyan,thesoviets} that the discontinuous jump in the extractable work from $W_{\rm indist}$ to $W_{\rm dist}$ is at odds with the fact that the similarity of any two gases can be varied smoothly\footnote{The statement that the similarity of two gases can be varied smoothly is best understood from the operational perspective of a chemist in the lab with a collection of gas specimens. The chemist can measure the properties of the gases (e.g. mass, condensation temperature, solubility etc.) to obtain quantitative measures of the difference between the gases. Considering the fundamental elementary composition of the gases one might question the degree to which such difference measures ultimately vary \textit{smoothly}. However, the point remains that while one can consider gases with varying degree of similarity, the work that can be extracted is discontinuous, depending only on whether the gases are identical or different.}. 
However, arguably~\cite{JaynesGibbs, SaundersGibbs, DieksGibbs} this tension is not especially mysterious because, while one can consider gases with varying degrees of similarity (e.g. in terms of composition or mass), it ultimately only matters whether or not two gases can be operationally distinguished.
It is important to emphasise here that the classical Gibbs mixing thought experiment applies to the mixing of \textit{homogeneous} gases, i.e. gases consisting of only particles of the same type. Inhomogeneous, i.e. gases consisting of different types of gas particles to start with, can be partially distinguishable. With the restriction to homogeneous gases, Gibbs' thought experiment highlights that fundamentally any two classical particles can either be distinguished or not, and any difficulty doing so is an epistemic limitation due to lack of knowledge and ability to identify suitable membranes on the part of the experimentalist.

However, for quantum systems this does not hold true. 
In the quantum setting two homogeneous gases could be distinguished not by their isotope or molecular composition but rather by their internal state. For example, a gas could consist of hydrogen atoms in their ground state or of hydrogen atoms in the first excited state. But it is also possible for each hydrogen  atom to be in a superposition of its ground state and excited state, and the distinguishability of such a superposition state with respect to the ground state can vary smoothly. In contrast to the classical case, one can thus consider two homogeneous gases, gas $R$ consisting of hydrogen atoms in the ground state and gas $L$ consisting of hydrogen atoms in an energetic superposition state, that are neither perfectly distinguishable nor perfectly indistinguishable. The quantum regime thus offers the ability to smoothly vary the distinguishability of two homogeneous gases and explore its impact on the thermodynamics of mixing.

\section{The Optomechanical Setup} \label{sec:opto}

\begin{figure}[t]
\centering
{\includegraphics[width=0.45\linewidth]{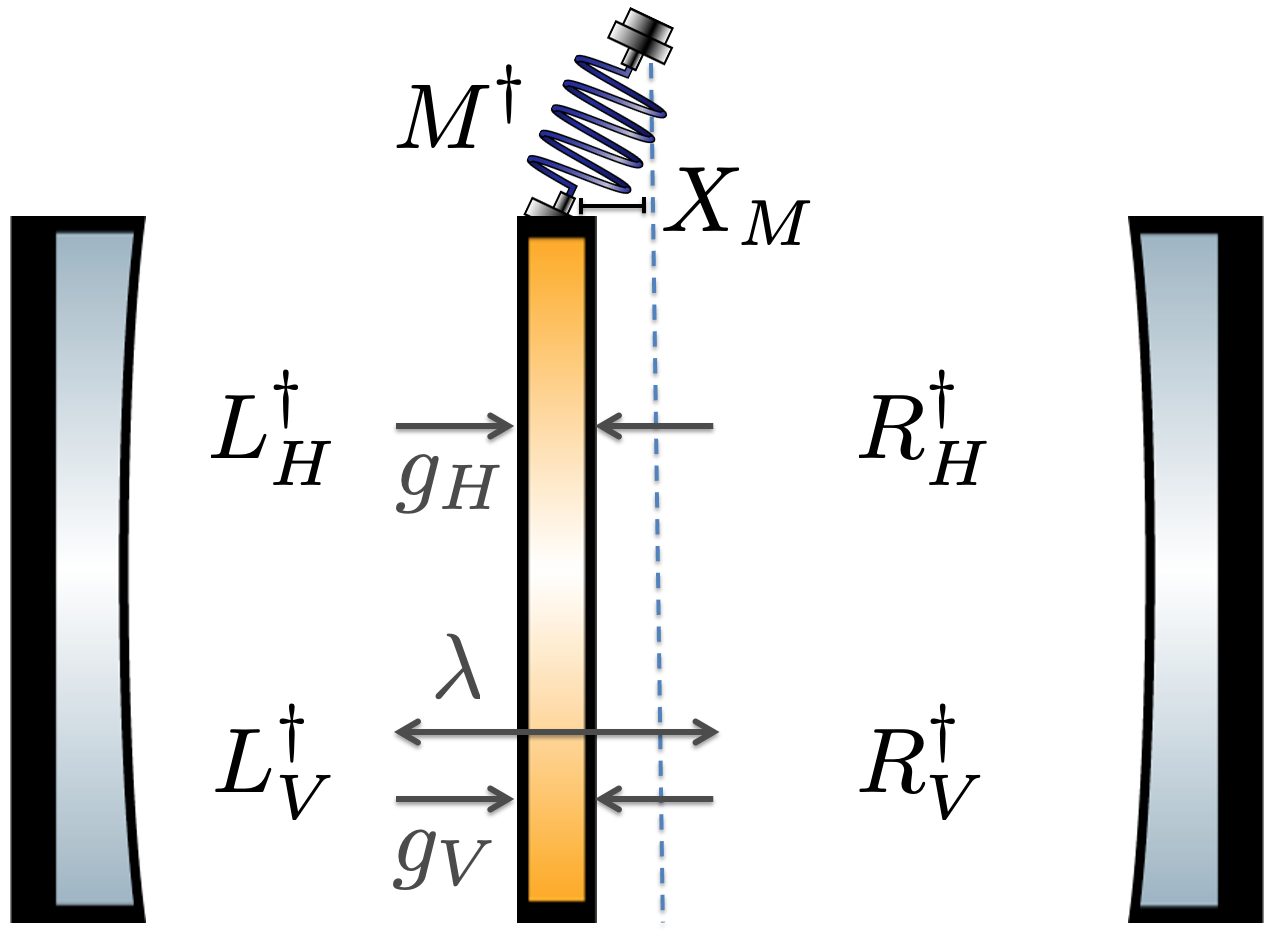}}
\caption{{\bf Optomechanical setup. }{A microscopically thin polarisation dependent beam-splitter membrane (PBS) is inserted into a cavity and allowed to oscillate. The displacement of the membrane from the mid-point of the cavity to the left is denoted by $X_M$ and the creation operator for this mechanical mode is $M^\dagger$. The transmission rate between the vertically polarised photon modes on the left hand side of the membrane, $L_V^\dagger$, and vertically polarised modes on the right, $R_V^\dagger$, is determined by the constant $\lambda$. The force exerted per horizontally (vertically) polarised photon on the membrane is given by $g_H$ ($g_V$). As the membrane reflects all horizontally polarised photons the force exerted on the membrane by horizontally polarised photons is greater than that of the vertically polarised photons, that is $g_H > g_V$}.}
\label{fig:optomechplot} 
\end{figure}

Here we introduce a physical setting that enables the study \zo{of \textit{quantum Gibbs mixing}, that is Gibbs mixing with homogeneous quantum gases that are partially distinguishable in virtue of occupying non-orthogonal internal quantum states~\cite{Schrodinger, thesoviets, Lande1, Lande2, allahverdyan,Peres,Maruyama}.}
In general, it is hard to conceive of realistic semi-permeable membranes that can differentiate between different internal states of an atom or molecule, such as the ground and excited state of a hydrogen atom. However, for photon gases of different polarisations~\cite{Peres}, a polarisation dependent beamsplitter satisfies all requirements.
The PBS membrane we consider acts as a mirror for horizontally polarised ($H$) photons, while acting as standard beamsplitter for vertically polarised ($V$) photons~\cite{zeilinger}. Thus similarly to a classical semi-permeable membrane the PBS is permeable to one gas ($V$) and impermeable to the other ($H$). However, unlike a classical semi-permeable membrane, a PBS acts coherently on the $V$ gas, generating superpositions of the left and right $V$ modes, thus generalising the semi-permeable membrane to the quantum realm.

One could conceive of generalising the classical Gibbs mixing protocol, sketched in Fig.~\ref{fig:classicalmixing}, by simply replacing the classical gases with photon gases and the semi-permeable membranes with a PBS; however, in that case the pistons are implicitly macroscopic and therefore the work output would remain entirely classical. Instead, we will here model the PBS membrane as a microscopic system such that the pistons, as well as the gases, are quantum mechanical. We further simplify the setting and consider a single piston membrane, as shown in Fig.~\ref{fig:optomechplot}, rather than the pair considered in the original classical protocol. This reduces the number of light modes and mechanical modes required, making both the calculations more tractable and the experimental setting more feasible. However, the core physics of the thought experiment remains unchanged. 

Specifically, we consider an optomechanical generalisation of the Gibbs mixing thought experiment in which, as sketched in Fig.~\ref{fig:optomechplot}, a PBS attached to a microscopic cantilever has been placed in the centre of an optical cavity. Gas $L$ (initially on the left of the cavity) consists of vertically polarised photons, each in state $\ket{V}$, and gas $R$ (initially on the right) consists of photons each polarised in a superposition state
\begin{equation}
    \ket{\theta} = \cos(\theta) \ket{V} + \sin(\theta) \ket{H} , 
\end{equation}
This allows us to explore the transition from perfectly indistinguishable to perfectly distinguishable gases by varying $\theta$ from $0$ to $\pi/2$.

\paragraph*{Hamiltonian.}
The Hamiltonian describing the `PBS-in-the-middle' optomechanical setup,
\begin{equation}\label{Eq: HamGen}
\begin{aligned}
    H&= H_0 +  H_{\mathrm{PBS}} +   H_{\mathrm{int}}  \, ,
\end{aligned}
\end{equation} 
is comprised of the non-interacting Hamiltonians of the photons and membrane, $H_0$, the polarisation dependent beamsplitter interaction between the photons in the left and right halves of the cavity, $ H_{\mathrm{PBS}}$, and the optomechanical interaction between the photons and the membrane, $ H_{\mathrm{int}}$~\cite{BunchingOptMech, experiment1a}. The Hamiltonian $H_0$ for the independent optical and mechanical modes can be written explicitly as
\begin{equation}
\begin{aligned}
H_0 = \, &\hbar \omega \left(  R_H^\dagger R_H  +  L_H^\dagger L_H  \right) + \hbar \omega \left(  R_V^\dagger R_V  +  L_V^\dagger L_V  \right) +  \hbar \omega_M M^\dagger M  \, .
\end{aligned}
\end{equation}
Here $\omega_M$ is the mechanical frequency of the membrane, $\omega$ is the light frequency in both halves of the cavity, $M^\dagger$ is the creation operator for the phonon membrane mode and $L_{H}^\dagger$ ($R_{H}^\dagger$) and $L_{V}^\dagger$ ($R_{V}^\dagger$) are the creation operators for horizontally and vertically polarised photons respectively in the left (right) cavity~\cite{polarisationOptomech}, for which the standard commutation relations hold, i.e. $[L_j , L_k^\dagger] = \delta_{i,j}$, $[R_j , L_k^\dagger] = 0$ etc. for $j, k= V, H$.

% While a standard BS, as considered in the standard `membrane-in-the-middle' setting~\cite{experiment1, BunchingOptMech}, transmits/reflects photons of any polarisation, \nn{the PBS considered here will only act as a standard BS on vertically polarised photons while acting as a mirror for horizontally polarised photons. 
While a standard BS, as considered in the standard `membrane-in-the-middle' setting~\cite{experiment1, BunchingOptMech}, transmits and reflects photons of any polarisation, the PBS considered here always reflects horizontally polarised photons. 
Therefore the coupling between the modes in the left and right halves of the cavity is of the form
\begin{equation}
\begin{aligned}
  H_{\mathrm{PBS}}  = \, & \frac{\hbar \lambda}{2} \left(R_V^\dagger L_V + L_V^\dagger R_V \right) \, ,
\end{aligned}
\end{equation} 
i.e. a vertically polarised photon on the left is annihilated ($L_V$) while a vertically polarised photon is created coherently on the right ($R_V^\dagger$), and vice versa, but there is no such coupling between the left and right modes for horizontally polarised photons\footnote{Note, if the membrane were perfectly (rather than partially) transmissive then the V modes would occupy the whole cavity and the left and right V modes of the cavity would not be well defined. Consequently, it would not be possible to define the initial unmixed state of the photon gases.}.
The coefficient $\lambda$ determines the reflection $r_V$ and transmission $t_V$ coefficients of the vertically polarised photons at time $t$, with the evolution of the photonic modes induced by $ H_{\mathrm{PBS}}  $ alone (in the Heisenberg picture) given by
\begin{equation}
\begin{pmatrix}
L_V(t) \\
R_V(t) \\
L_H(t) \\
R_H(t) \\  
\end{pmatrix} = U_{\rm PBS}
\begin{pmatrix}
L_V(0) \\
R_V(0) \\
L_H(0) \\
R_H(0) \\  
\end{pmatrix}
\ \  \text{with} \ \ U_{\rm PBS} = \begin{pmatrix}
r_V & -i t_V & 0 & 0\\
-i t_V & r_V & 0 & 0 \\
0 & 0 & 1 & 0 \\
0 & 0 & 0 & 1  
\end{pmatrix} \, ,
\end{equation}
where $r_{V} (t) = \cos\left(\frac{\lambda}{2} t\right)$ and $t_V (t)= \sin\left(\frac{\lambda}{2} t\right)$~\cite{zeilinger}.  The horizontally polarised photons in the left and right halves of the cavity are confined to their respective halves and thus are uncoupled such that transmission coefficient for horizontally polarised photons vanishes, $t_H=0$, and the photons are always reflected, $r_H=1$. 
The key quantum effect is that a single photon initially in a superposition state $\ket{\theta}$ will in time interact with both the beamsplitter and mirror components of the PBS. 

% \red{Note: we have now said three times in the paper (+ once in a footnote) that the "PBS considered here will only act as a standard BS on vertically polarised photons while acting as a mirror for horizontally polarised photons." -- maybe we should cut at least once?}

When photons collide with the membrane they will exchange momentum and exert a pressure (radiation pressure) on the surface of the membrane~\cite{RadiationPressure} that depends on whether a photon is reflected or transmitted. The total force exerted by photons on the membrane,
\begin{equation}\label{Eq:TotalForce}
F(g_H, g_V) = \hbar g_H \Delta N_{H}
+ \hbar g_V \Delta N_{V} \, ,
\end{equation}
is the sum of the products of the forces exerted per photon, $\hbar g_H$  ($ \hbar g_V$) for a horizontally (vertically) polarised photon, and the difference between the numbers of photons in the left and right halves of the cavity, $\Delta N_{H} := L_H^\dagger L_H - R_H^\dagger R_H$  (and $\Delta N_{V} := L_V^\dagger L_V - R_V^\dagger R_V $) for horizontally (and vertically) polarised photons. In general $g_V$ and $g_H$ are independent variables; however, since all horizontal photons are reflected but vertical photons are partially reflected and partially transmitted, the force exerted by horizontal photons on the membrane should be greater than that of vertical photons which implies that $g_H$ is larger than $g_V$.
This radiation pressure gives rise to an optomechanical interaction energy of the form
\begin{align}\label{eq:OptMechH}
& H_{\mathrm{int}}  (g_H, g_V) = - F(g_H, g_V) X_M \; ,
\end{align}
where $X_M= x_{\rm zpf}( M+M^\dagger)$ is the displacement of the membrane from the centre of the cavity. The prefactor $x_{\rm  zpf}$ is the mechanical oscillator's zero point uncertainty $x_{\rm zpf} = \sqrt{\hbar / 2 m \omega_M}$ with $m$ the mass of the membrane.

\paragraph*{Initial state.}
While in the classical setting the number of particles in each gas is constant and the gases are in thermal contact with a heat bath such that their temperature is also fixed, here the photon gases do not thermalise through mutual interactions with a heat bath and the gases cannot have \textit{both} a fixed temperature and photon number. 
However, the cases of fixed temperature and fixed photon number can be studied separately. For the initial state of our photon gases we first consider a Fock state configuration which has precisely $n$ photons per cavity (but no notion of temperature) and thus the initial state of the photons can be written as $\ket{\psi_{\rm F}^n} := \ket{\psi_L^n (0)}\otimes  \ket{\psi_R^n (\theta)}$ with
\begin{equation}
\begin{aligned}\label{eq: fockstate}
&\ket{\psi_C^n (\theta)} \propto \left(\cos(\theta) \, C_V^\dagger + \sin(\theta) \, C_H^\dagger\right)^{n} \ket{0}  \ \ \ \text{for} \ \ \ C = R, L \; .
\end{aligned}    
\end{equation}
Secondly, we consider a photon gas $\rho^T := \gamma_{L}^T (0) \otimes \gamma_{R}^T (\theta)$ in which the photons in each cavity are initially in a thermal state at temperature $T$, with 
\begin{equation}\label{eq: Thermalphotonstates}
\begin{aligned}
&\gamma_{C}^T (\theta) \propto \sum_{n=0}^{\infty} e^{- {n \hbar \omega/ k_B T}} \ket{\psi_C^n (\theta)} \bra{\psi_C^n ( \theta)} \ \ \ \text{for} \ \ \ C = R, L .
\end{aligned}
\end{equation}
This configuration has a fixed temperature, but no fixed photon number per cavity. While the thermal configuration perhaps makes better contact with the thermodynamics which is central to the original Gibbs mixing thought experiment, the Fock state configuration is conceptually interesting in virtue of the fact that the Fock state is a genuinely quantum mechanical state of light. 

In the usual classical setting, the states of the semi-permeable membranes are not explicitly modelled. However, presumably, the membranes are initially at equilibrium with the surrounding heat bath. On this basis, we will similarly take the membrane to be prepared in a thermal state,
\begin{equation} \label{eq:membranestate}
    \gamma_{M}^T  \propto \exp\left(- \frac{ \hbar \omega_M M^\dagger M}{k_B T}\right) \; ,
\end{equation}
where, in the thermal configuration, $T$ is also the temperature of the gases so that the setup is at thermal equilibrium.

\section{The Dynamics of Mixing}\label{sec:Calc}

Before describing how the initial photon gases evolve dynamically, let us briefly recap the situation. There are three key degrees of freedom of the photonic state: the polarisation degree of freedom, the spatial degree of freedom and photon number.
The state of the photons can be written in the number basis of the four modes,
\begin{equation} \label{eq:totH}
   \ket{\vec{n}} = \ket{n_{LH},n_{RH},n_{LV},n_{RV}} \propto  (R_H^\dagger)^{n_{RH}} (L_H^\dagger)^{n_{LH}} (R_V^\dagger)^{n_{RV}} (L_V^\dagger)^{n_{LV}} \ket{0}
\end{equation}
with $n_{LH}$ photons in the left H mode, $n_{RH}$ photons in the right H mode, $n_{LV}$ photons in the left V mode, and $n_{RV}$ photons in the right V mode. The total state of the photons at any moment in time is a mixed state of the form 
\begin{equation}
\begin{aligned}
&\rho = \sum_{\vec{n},\vec{m}} \alpha(\vec{n},\vec{m})  \ket{\vec{n}} \bra{\vec{m}} \, .
\end{aligned}    
\end{equation}
The coefficients $ \alpha(\vec{n},\vec{m})$ are at first determined by the initial conditions (as detailed in the previous section) but then will evolve in time. In general, the dynamics of the photons will be rather complex with coherence generated with respect to the spatial degree of freedom.

An intuitive picture of the dynamics of the photons and membrane can be built without explicit calculation of the dynamics.
As $\rho_{\rm F}^n = \ket{\psi_{\rm F}^n}\bra{\psi_{\rm F}^n}$ and $\rho^T$ evolve under the total Hamiltonian \eqref{eq:totH}, horizontally polarised photons are confined to the right half of the cavity while vertically polarised photons are free to oscillate between the two halves, leading to a partial mixing of the gases.
When the two gases are perfectly distinguishable, i.e. $\theta = \pi/2$, the horizontally polarised photons, which are confined solely in the right half of the cavity, generate a net leftwards force on the PBS membrane and this displaces the membrane from the centre to the left. 
In contrast, when the two gases are indistinguishable, i.e. $\theta = 0$, both gases are vertically polarised and, as on average there will be equal numbers of photons in both halves of the cavity, there will be no net force on the membrane. For indistinguishable gases we thus expect the membrane to remain on average in the centre of the cavity.
For partially distinguishable gases, i.e. $0<\theta<\pi/2$, the force exerted by the gases on the membrane, and therefore the membrane's displacement, is expected to continuously increase with increasing $\theta$.

In analogy to the classical mixing scenario, the displacement of the membrane can be associated with a work output, which is expected to increase with the distinguishability of the two gases. In contrast to the classical case discussed in Section~\ref{sec:Gibbs}, where the thermal bath replenishes the energy lost by the photon gases to the membrane such that the work done on the piston ultimately stems from the heat bath, here the system is closed and therefore the work done on the piston membrane stems solely from the effective energy of the photon gases.

To quantitatively study the dynamics of the light field and mechanical degree of freedom it is helpful to work in the Heisenberg picture since this makes it possible to obtain general results for the evolution of the relevant observables, that is the membrane's displacement and energy, for a general initial state of the photon gases and membrane. To maintain an analogy with classical Gibbs mixing, where the number of gas molecules is fixed and the motion of the pistons are assumed to be frictionless, we here consider an idealised cavity and membrane which experience no dissipative effects. This assumption is relaxed in Section~\ref{sec:GibbsExp} where more realistic experimental implementations are discussed.

The membrane evolves as a quantum harmonic oscillator driven by the radiation pressure from the photons in the cavity. Working in the Heisenberg picture, the equation of motion of the membrane reads 
\begin{equation}\label{Eq:MotionMembrane}
    \frac{d^2 X_M}{d t^2} + \omega_M^2 X_M = \frac{F(g_H, g_V, t)}{m} \ ,
\end{equation}
where the time dependence of the total force exerted by the photons on the membrane, Eq.~\eqref{Eq:TotalForce}, is explicitly included.
Since the PBS membrane acts as a mirror for horizontally polarised photons, the number of horizontally polarised photons in the right (and hence left) cavity is conserved for evolutions under the total Hamiltonian, and therefore 
\begin{equation}\label{Eq:MotionVerticalPhotons}
\Delta N_{H}(t) = \Delta N_{H}(0) \ . 
\end{equation}
However, the vertically polarised photons are free to oscillate between the two halves of the cavity with their number difference $\Delta N_V$ coupled to the position of the membrane via
\begin{align}\label{Eq:DeltaNHdiff}
    &\frac{d^2 \Delta N_V}{dt^2} = -  \lambda^2 \Delta N_V - 2 g_V \lambda X_M (L_V^\dagger R_V + L_V R_V^\dagger) \ . 
\end{align}
Note, that the total Hamiltonian commutes with $L_V^\dagger L_V + R_V^\dagger R_V$ and $L_H^\dagger L_H + R_H^\dagger R_H$ and therefore the total number of horizontally and vertically polarised photons is conserved.

The dynamics of the photons and membrane are thus determined by the coupled differential equations for the light and mechanical modes, Eq.~\eqref{Eq:MotionMembrane} and Eq.~\eqref{Eq:DeltaNHdiff}. While solving these exactly is prohibitively difficult, a perturbative solution that describes the dynamics of the expectation of selected observables can be constructed. Given that the force exerted by a single photon on the massive membrane is weak\footnote{Note, we are implicitly still considering what is usually called the strong coupling regime here in that we have assumed that the cavity damping rate $\kappa$ is sufficiently small that we can disregard it and therefore $g_H x_{\rm zpf} > \kappa$ and $g_V x_{\rm zpf} > \kappa$.}, the single photon coupling strengths, $g_H x_{\rm zpf}$ and $g_V x_{\rm  zpf}$, are expected to be small compared to the PBS coupling strength, $\lambda$. We thus choose to solve the dynamics perturbatively in $g_H$ and $g_V$.

As the force exerted on the membrane is already linear in $g_H$ and $g_V$, Eq.~\eqref{Eq:TotalForce}, studying the dynamics of the membrane to 1st order in $g_H$ and $g_V$ amounts to disregarding the back-action that the motion of the membrane has on the dynamics of the photons and evaluating the dynamics of the photons to 0th order in $g_H$ and $g_V$. It follows from Eq.~\eqref{Eq:DeltaNHdiff} that, in the absence of back-action, the vertically polarised photons oscillate at a rate $\lambda$ between the two halves of the cavity according to
\begin{equation}\label{Eq:MotionHorizontalPhotons1stOrder}
    \Delta N_V(t) =  \Delta N_V (0) \cos(  \lambda t ) + \Delta K_V (0) \sin( \lambda t) \ , 
\end{equation}
where
\begin{equation}
    \Delta K_V := i  ( R_V^\dagger L_V -  R_V L_V^\dagger ) \ .
\end{equation}

The membrane dynamics \eqref{Eq:MotionMembrane} is thus driven by the sum of an oscillatory force, originating from the oscillatory motion of the vertical photons between the left and right halves of the cavity, and a constant force, originating from the initial imbalance $\Delta N_H (0) \neq 0$ of the horizontal photons.
The solution to the equation of motion of the membrane, 
\begin{equation}\label{eq:membranepositionop}
\begin{aligned}
X_M(t) =  \frac{ \hbar g_H }{ m \omega_M^2} \, \Delta N_H(0) + X_M^{\rm osc}(t)  \, ,
\end{aligned}
\end{equation}
is composed of a displacement proportional to the difference in number of horizontal polarised photons in the left and right halves of the cavity and an oscillatory term, $X_M^{\rm  osc}(t)$. This oscillatory term is of the form 
\begin{equation}\label{eq:Xosc}
\begin{aligned}
    X_M^{\rm osc}(t) 
    = &   v_V(t)  \, \frac{\hbar g_V}{m \omega_M^2}  \Delta K_V(0) + \frac{\dot{v}_V(t)}{\lambda} \frac{\hbar g_V}{m \omega_M^2}  \Delta N_V(0) \\ 
    &+ \cos(\omega_M t) \left(X_M(0) - \frac{\hbar g_H}{m \omega_M^2}  \Delta N_H(0) \right) + \sin(\omega_M t) \, \frac{P_M(0)}{m \omega_M}  \, ,
\end{aligned}
\end{equation}
where $P_M(0)$ is the membrane's momentum operator in Heisenberg picture at time $t=0$ and $v_V(t)= \frac{\omega_M^2}{\omega_M^2 - \lambda^2}\left(\sin(\lambda t)- \frac{\lambda}{\omega_M}\sin(\omega_M t) \right)$ is an oscillatory function quantifying beating between the driving frequency, $\lambda$, and the membrane's natural frequency, $\omega_M$. 
Having found the membrane's dynamics~\eqref{eq:membranepositionop} with oscillatory term \eqref{eq:Xosc},
we are now ready to quantify the thermodynamics of mixing within the present optomechanical setting.

\section{Work output from the Gibbs mixing of photons}\label{sec:OptMechMixResults}

To compare the work done by the mixing of the photon gases to that in Gibbs' classical thought experiment, one needs to introduce a measure of work done on the membrane. The question of how to define work and heat in the quantum regime, where energetic coherences can be present, has been discussed extensively elsewhere~\cite{defofquantumwork1,defofquantumwork2,defofquantumwork3,defofquantumwork3.5,defofquantumwork4,defofquantumwork5}. Here we sidestep these fundamental difficulties, and identify a natural candidate for the work output in the present optomechanical setting by drawing an analogy with classical Gibbs mixing.

In the classical Gibbs mixing protocol described in Fig.~\ref{fig:classicalmixing}, one could imagine the work done by the expanding gases is stored by the pair of springs that are compressed during the mixing process. By analogy, one might consider the work done on the PBS membrane to be the potential energy associated with its displacement resulting from the mixing of the  photon gases, $W_M (t) \propto \langle X_M (t) \rangle^2$.
However, in contrast to the classical protocol where the final displacement of the membrane is constant and well defined, in the optomechanical setting the membrane oscillates about its new displaced origin in time and there is some spread to its wave function. In order to have a well defined work output, we therefore take the work done from mixing as given by the potential energy associated with the oscillator's {\it time averaged} displaced origin. That is
\begin{equation}\label{eq:DefOfWork1}
    W^{\rm mix}_M := \frac{1}{2} m \omega_M^2 \langle \bar{X}_M \rangle^2 \; ,
\end{equation} 
where
\begin{equation}
  \begin{aligned}
 \langle \bar{X}_M \rangle  :=  \Tr \left[  \frac{1}{\tau }\int_0^{\tau}dt X_M(t) \, \, \rho \right] \ ,  
\end{aligned}\label{eq:TimeAverage}
\end{equation}
is the displacement of the membrane averaged over both an oscillation cycle of time $\tau$ and the initial quantum state of the two photon gases and the membrane, i.e. either $\rho = \rho_{\rm F}^n \otimes \gamma_{M}^T$ for Fock state photons or $\rho =\rho^T \otimes \gamma_{M}^T$ for thermal state photons, see \eqref{eq: Thermalphotonstates} and \eqref{eq:membranestate}.

\begin{figure}[t]
\centering
{\includegraphics[width=0.95\linewidth]{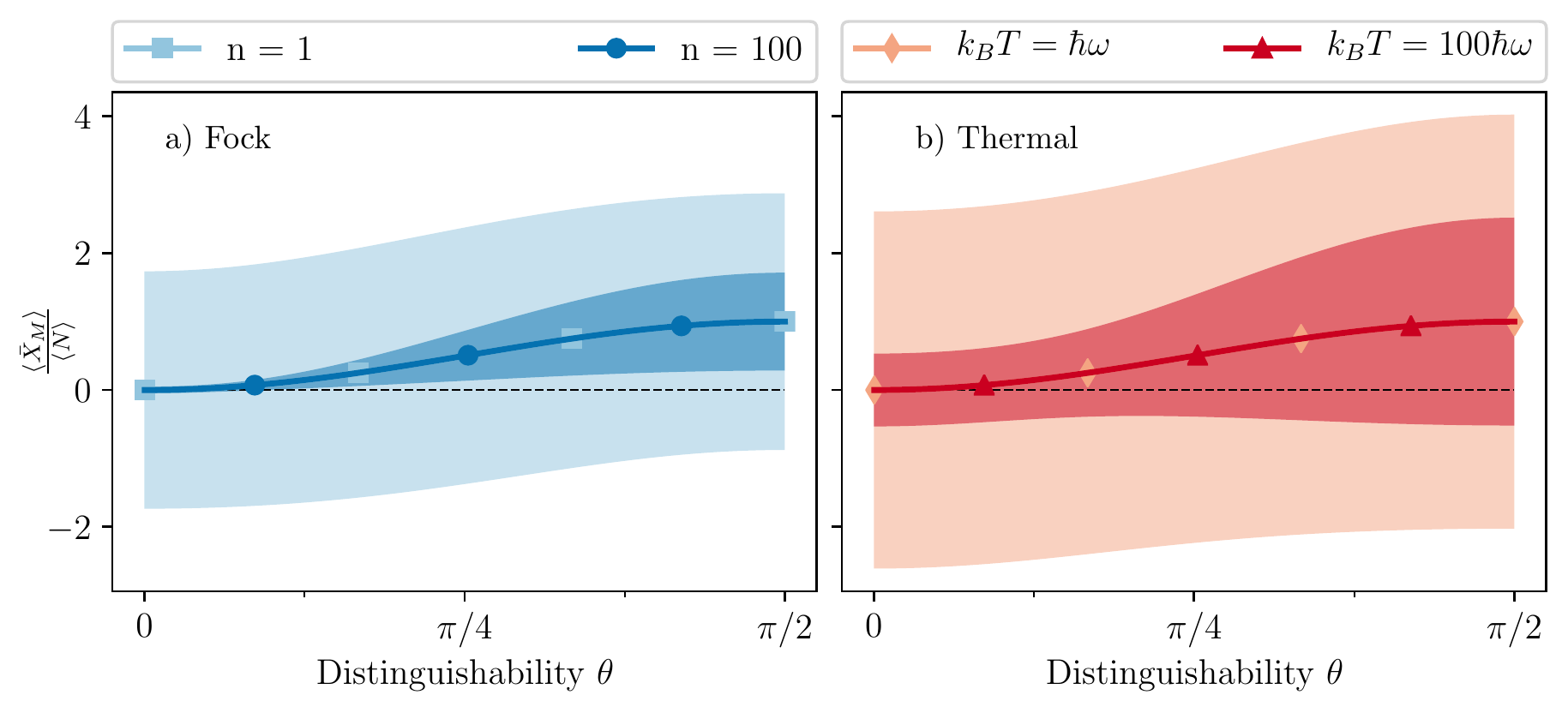}} 
\caption{\label{fig:meandisplacementPBS} 
{\bf Time averaged displacement of PBS membrane.} Plot of $\langle \bar{X}_M \rangle$ (directly related to the work output via \eqref{eq: W1to2ndOrder}) normalised by the mean number of photons $\langle N \rangle$ per gas as a function of the distinguishability $\theta$ of the photon gases.
{\bf a)} Photons are initially in a Fock state $\rho_F^n$ with $n=1$ photons per cavity (light blue squares) and $n=100$ photons per cavity (dark blue circles), while the membrane starts in a thermal state at temperature $T = \omega_M/k_B$ for both cases. 
{\bf b)} Photons and membrane are both initially in a thermal state $\rho^T$ at temperature $T = \hbar \omega /k_B = 10 \hbar \omega_M/k_B$ (pink diamonds) and $T = 100 \hbar \omega/k_B = 1000 \hbar \omega_M/k_B$ (red triangles).
For all four cases (light blue squares, dark blue circles, red triangles, pink diamonds) the displacement per photon is the same function over the distinguishability parameter $\theta$. The shaded region shows the normalised standard deviation of the displacement, $\frac{\sqrt{\langle \bar{X}_M^2 \rangle - \langle \bar{X}_M \rangle^2}}{\langle N \rangle}$,
showing decreasing membrane displacement uncertainty per photon for higher number Fock states and higher temperature thermal states, as expected.
In both plots the following parameters were chosen to illustrate the thought experiment: $\lambda/\omega_M  = 2$, $g_H/g_V = 6$, $\omega_M /g_V x_{zpf} = 10$ and $\omega/\omega_M = 10$. All distances are given in units of $\frac{\hbar g_H}{m \omega_M^2}$.}
\end{figure}

The work that can be extracted from the mixing of partially distinguishable photon gases can now be calculated from the expression for the dynamics of the membrane in the Heisenberg picture, Eq.~\eqref{eq:membranepositionop}. Since the oscillatory component to the membrane's motion  $X_M^{\rm osc}(t)$ vanishes when averaged over a complete oscillation cycle, the time averaged displacement of the membrane,
\begin{equation}
   \begin{aligned}
  \langle  \bar{X}_M \rangle =  \frac{ \hbar g_H }{ m \omega_M^2} \langle \Delta N_H(0) \rangle \ ,
\end{aligned}
\end{equation}
is directly proportional to the difference $\langle \Delta N_H(0) \rangle$ in the initial number of horizontally polarised photons in the two gases.
For the initial photon gas states $\rho_{\rm F}^n$ or $\rho^T$, both gases have the same total number of photons $\langle N \rangle = \langle C_H^\dagger(0) C_H(0) +  C_V^\dagger(0) C_V(0) \rangle $ per gas $C=L, R$. I.e. $\langle N \rangle =n$ per gas for the Fock states \eqref{eq: fockstate} and $\langle N \rangle= 1/(e^{\hbar \omega/k_B T}-1)$ per gas for  thermal states \eqref{eq: Thermalphotonstates}.
However, the number of {\it horizontally} polarised photons in the right gas is $\langle N \rangle \sin^2\theta$ while on the left it is zero.
Hence, the time averaged displacement of the membrane is
\begin{equation}\label{eq:TimeAverageDisplacement}
   \begin{aligned}
\langle  \bar{X}_M \rangle=  \frac{\hbar  g_H  \langle N \rangle}{ \omega_M^2 m}\sin^2(\theta)  \ .
\end{aligned}
\end{equation}
It thus follows that the potential energy associated with the average displacement of the membrane due to mixing of the two photon gases is
\begin{equation}\label{eq: W1to2ndOrder}
     W_M^{\rm mix} = \frac{\hbar^2 g_H^2 \langle N \rangle^2}{ 2 \omega_M^2 m} \sin^4(\theta) \ .
\end{equation} 
This is the mixing work associated with the mixing of two homogeneous quantum gases in the "PBS-in-the-middle" setting. The expression implies that the work output vanishes for perfectly indistinguishable gases and is maximised for perfectly distinguishable gases, in agreement with the classical case, as one would expect. However, crucially, in contrast to classical Gibbs mixing and inline with prior analyses, e.g. in terms of the mixing entropy by Luboshitz and Podgoretskii~\cite{thesoviets} and the ergotropy by Allahverdyan and Nieuwenhuizen~\cite{ergotropy,allahverdyan}, the work $ W_M^{\rm mix} \propto \langle\bar{X}_M\rangle^2$ now \textit{smoothly} increases between these two extremes. See  Fig.~\ref{fig:meandisplacementPBS} for a plot of $\langle\bar{X}_M\rangle$ and its standard deviation for the Fock and thermal photon gas states.

The mixing work is highly general in the sense that it holds not only for the initial Fock and thermal photon gas states, but also for any initial photon gas states that have the same number distribution while the left gas is $V$-polarised and the right gas is $\theta$-polarised.
As such, the work output is independent of the purity of the number distribution of the two quantum gases. That is, any coherence in the number basis has no effect on the work output. 

\medskip

One could also consider mixing inhomogeneous gases, which in the optomechanical setting considered here, would correspond to mixing one gas initially in the state $\rho_R(\theta) = \cos^2(\theta) | V \rangle \langle V| + \sin^2(\theta)| H \rangle \langle H| $ and the other initially in the state $\rho_L(0) =| V \rangle \langle V|$. 
Since Eq.~\eqref{eq: W1to2ndOrder}, is independent of coherence with respect to the polarisation degree of freedom, it also gives the work that can be extracted from mixing two such gases. 
However, as we remarked in Section~\ref{sec:Gibbs}, mixing \textit{inhomogeneous} gases would deviate from the spirit of Gibbs' original thought-experiment concerning the mixing of {\it homogeneous} gases. Conceptually, the mixing of partially distinguishable inhomogeneous classical gases is fundamentally different from the mixing of partially distinguishable quantum gases in virtue of the fundamental distinction between classical mixtures and quantum superpositions. Crucially, restricting one's attention to homogeneous gases, only quantum gases can be partially distinguishable. Our proposed implementation of quantum Gibbs mixing highlights this simple but key difference in the nature of distinguishability in quantum and classical physics.

\section{Optomechanical Gibbs mixing analysis}\label{sec:fluctuations}

Before proceeding to discuss how one might observe the continuous work output with distinguishability experimentally in Section~\ref{sec:GibbsExp}, here we provide a more detailed account of the thermodynamic properties of Gibbs mixing in this optomechanical setting.

\paragraph{Fluctuations in the work output.}
A key distinction between Gibbs mixing in the optomechanical setting compared to prior studies concerns the fluctuations in the work output. When the pistons are assumed to be macroscopic classical systems, as in Gibbs' classical analysis~\cite{Gibbs} and in prior quantum analyses based on entropic arguments~\cite{Schrodinger,thesoviets,Lande1,Lande2}, the fluctuations in the work output from mixing are negligible. Here, however, as shown in Fig.~\ref{fig:meandisplacementPBS}, the standard deviation of the displacement of the membrane is large as compared to the membrane's average displacement which implies that there are large fluctuations in $W^{\rm mix}_M$. 

These substantial fluctuations in the work output are a feature arising from using a microscopic quantum piston membrane that is sensitive to the microscopic dynamics of the photon gases. The fluctuations capture both the motion of the mechanical degree of freedom due to the oscillatory force exerted by the photons as well as its quantum uncertainty. This quantum component to the uncertainty stems from the initial vacuum fluctuations of the photonic and membrane modes and the coherent nature of the dynamics. In particular, the PBS interaction gives the photons access to coherent superpositions between the two halves of the cavity and the optomechanical interaction entangles the photonic modes and the microscopic piston generating quantum fluctuations in the membrane's position. 
As shown in light blue and pink in Fig.~\ref{fig:meandisplacementPBS}, the fluctuations per photon are largest for low photon numbers and low temperatures. This can be attributed to the increased relative significance of vacuum fluctuations when the number of photons in the cavity is small. 

\begin{figure}[t]
\centering
\includegraphics[width=0.6\linewidth]{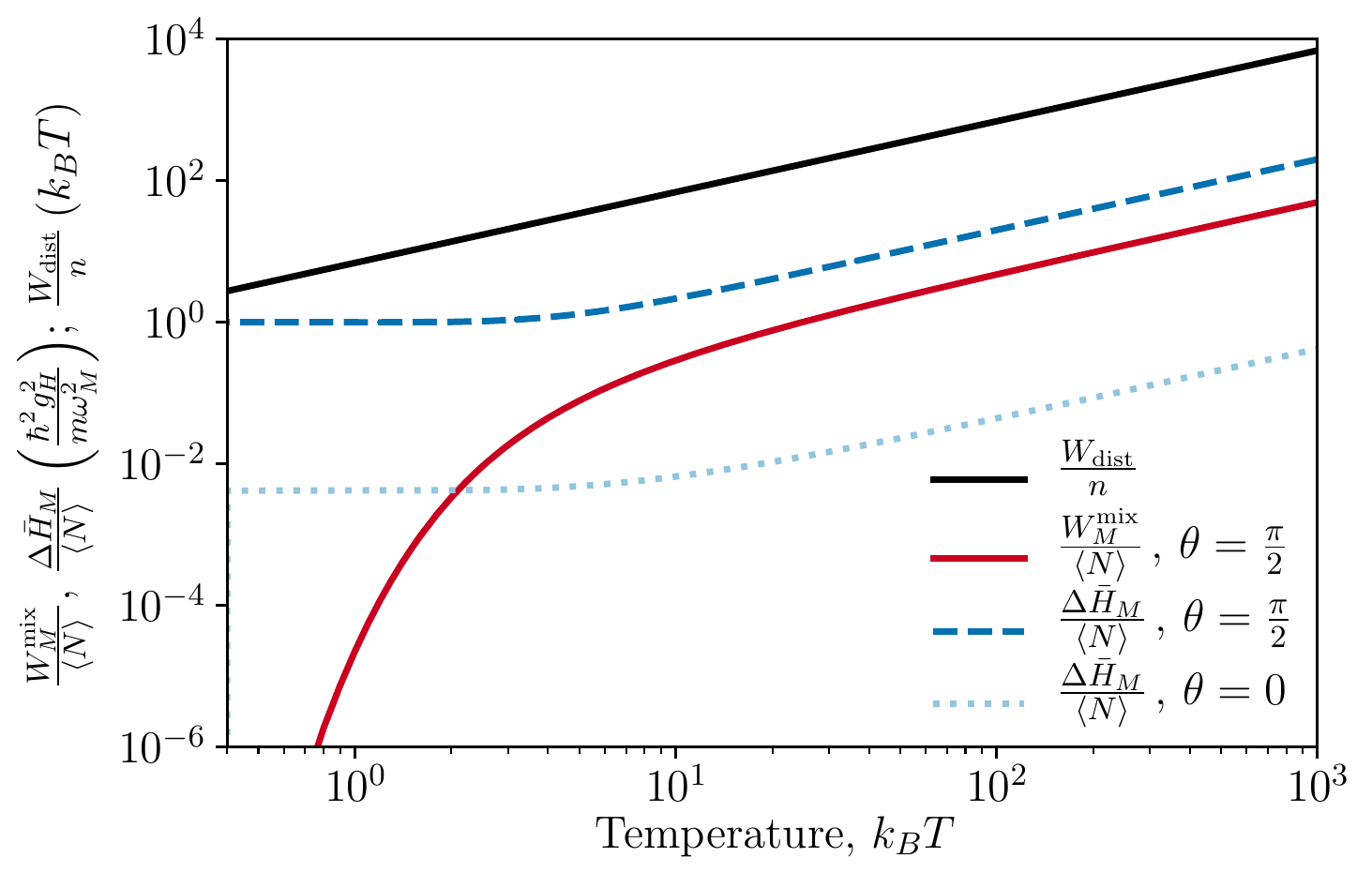}
\caption{\label{fig:tempdependence} 
{\bf Mixing work per particle done on the PBS membrane as a function of photon gas temperature.} For perfectly distinguishable gases ($\theta =\pi/2$, solid red) the mixing work $W^{\rm mix}_M/\langle N  \rangle$ (Eq.~\eqref{eq: W1to2ndOrder}, to first order in $g_V$ and $g_H$, normalised by the average photon gas number $\langle N \rangle$), grows as a function of the photon state temperature $T$, Eq.~\eqref{eq: Thermalphotonstates}, and is zero for indistinguishable gases ($\theta =0$, not shown). For higher temperatures the slope becomes the same as that of the classical Gibbs mixing work per particle  $W_{\rm dist}/n$  (solid black) for distinguishable gases, see Eq.~\eqref{eq:distingwork}. Also shown is the full energy transfer to the membrane given by $\Delta \bar{H}_M$ in Eq.~\eqref{eq:MembraneFluctTimeAverage} for perfectly distinguishable gases (dark blue dashed) and for perfectly indistinguishable gases (light blue dotted). Again in the high temperature regime, both slopes become the same as the classical mixing work (solid black) and hence the energy transfers to the membrane grows linearly with the photon gas thermal state temperature. 
The classical work is given in units of $k_B T$, and the quantum work $W_M$ and energy transfer $\Delta \bar{H}_M$ are given in units of $\frac{\hbar^2 g_H^2}{m \omega_M^2}$. In this plot $\lambda/\omega_M  = 2$, $g_H/g_V = 6$, $ \omega = 10 \omega_M$.
}
\end{figure}

\paragraph{Temperature dependence.}
We have until this point focused on the distinguishability dependence of the dynamics; however, classically the temperature dependence of the work output is pertinent and therefore it is natural to also discuss this dependence here. In the current setting, any temperature dependence of the work output $W^{\rm mix}_M$ enters through the initial state of the photons. Therefore the work output does not depend on temperature if the number of photons in each gas is temperature independent, such as in the Fock state. However, in the thermal state the average number of photons is $\langle N \rangle= 1/(e^{\hbar \omega/k_B T}-1)$ and its variance is $\langle (N-\langle N \rangle)^2 \rangle= e^{\hbar \omega/k_B T}/(e^{\hbar \omega/k_B T}-1)^2$.
The resulting temperature dependence of the work output is shown in Fig.~\ref{fig:tempdependence}. To aid comparison with classical Gibbs mixing work for two distinguishable gases, $W_{\rm dist} = 2 n \, k_B T \ln 2$ in Eq.~\eqref{eq:distingwork}, we plot the work done on the membrane \textit{per photon}, $W^{\rm mix}_M /\langle N \rangle$, to compensate for the fact that classically the number of particles in the gas is an arbitrary constant $n$, whereas for thermal photon gases the particle number $\langle N \rangle$ is temperature dependent. 
The log-log plot of work over temperature in  Fig.~\ref{fig:tempdependence} shows that the work output for both distinguishable and indistinguishable gases has the same slope as the classical work in the high temperature limit. Hence, since the classical mixing work is linear in $T$, the quantum mixing work is also linear in $T$ at high temperatures, in agreement with prior results~\cite{thesoviets,allahverdyan}.
For low temperatures, the work per photon $W^{\rm mix}_M /\langle N \rangle$ tends to zero at a super-linear rate, see Fig.~\ref{fig:tempdependence}. These low temperature deviations from the linear scaling are unsurprising given that the classical and entropic analyses are formulated for the asymptotic limit of large numbers of particles~\cite{thesoviets,Schrodinger,Lande1,Lande2}, whereas in the low temperature limit here, the gases consist of only a handful of photons. Moreover, whereas prior analyses~\cite{thesoviets,Schrodinger,Lande1,Lande2} quantify the maximal work that can be extracted from mixing, we have presented a protocol which we have no reason to expect to be optimal. Indeed the gases, only partially, rather than fully mix within our setting. This could also account for the diverging predictions. 

\paragraph{Total energy transfer.} In addition to the work output, another natural thermodynamic quantity to consider is the total energy transfer from the two photon gases to the membrane. In contrast to the work output, Eq.~\eqref{eq:DefOfWork1}, which does not account for the energy associated with the motion of the membrane, the total energy of the membrane is composed of a kinetic as well as a potential energy component. Thus, while the work output derives entirely from the constant force exerted by the horizontally polarised photons, the driving force provided by the oscillatory motion of the photons also contributes to the membrane's energy. Since the dynamics generates oscillatory behaviour and entangles the mechanical and optical degrees of freedom, it is most insightful to consider the time-average of the quantum mechanical expectation value of change in the energy of the membrane,
\begin{equation}\label{eq:DefOfWork2}
    \Delta \bar{H}_M := \frac{1}{\tau }\int_0^{\tau} \Tr[ (H_M(t) - H_M(0)) \rho ] \, dt  \ ,
\end{equation}
where $H_M = \hbar \omega_M M^\dagger M$ is the Hamiltonian of the membrane and the time average is taken over a complete oscillation cycle, as in Eq.~\eqref{eq:TimeAverage}. Using the expression for the evolution of the membrane position operator, Eq.~\eqref{eq:membranepositionop}, and its derivative,
the energy transfer to the membrane is found to be of the form
\begin{equation}\label{eq:MembraneFluctTimeAverage}
    \Delta \bar{H}_M  =  \alpha \langle  \Delta N_H^2(0) \rangle +  \beta \langle \Delta N_H(0) \Delta N_V(0)  \rangle + \eta \langle \Delta N_V^2(0) \rangle + \mu \langle  \Delta K_V^2(0) \rangle  \; ,
\end{equation}
where the prefactors $\alpha = \frac{\hbar^2 g_H^2  }{ m \omega_M^2} $, $\beta = \frac{\hbar^2 g_H g_V}{ m (\omega_M^2 -  \lambda^2 ) }$, $\mu =  \frac{\hbar^2 g_V^2 (3 \lambda^2 + \omega_M^2)}{4 m ( \lambda^2 - \omega_M^2)^2} $ and $\eta =\frac{\hbar^2 g_V^2 ( \lambda^2 + 3 \omega_M^2)}{4 m ( \lambda^2 - \omega_M^2)^2}$ depend only on system parameters but not on the initial state of the gases or membrane.

\begin{figure}[t]
\centering
\subfloat[Energy transfer to PBS]{\includegraphics[width=0.475\linewidth]{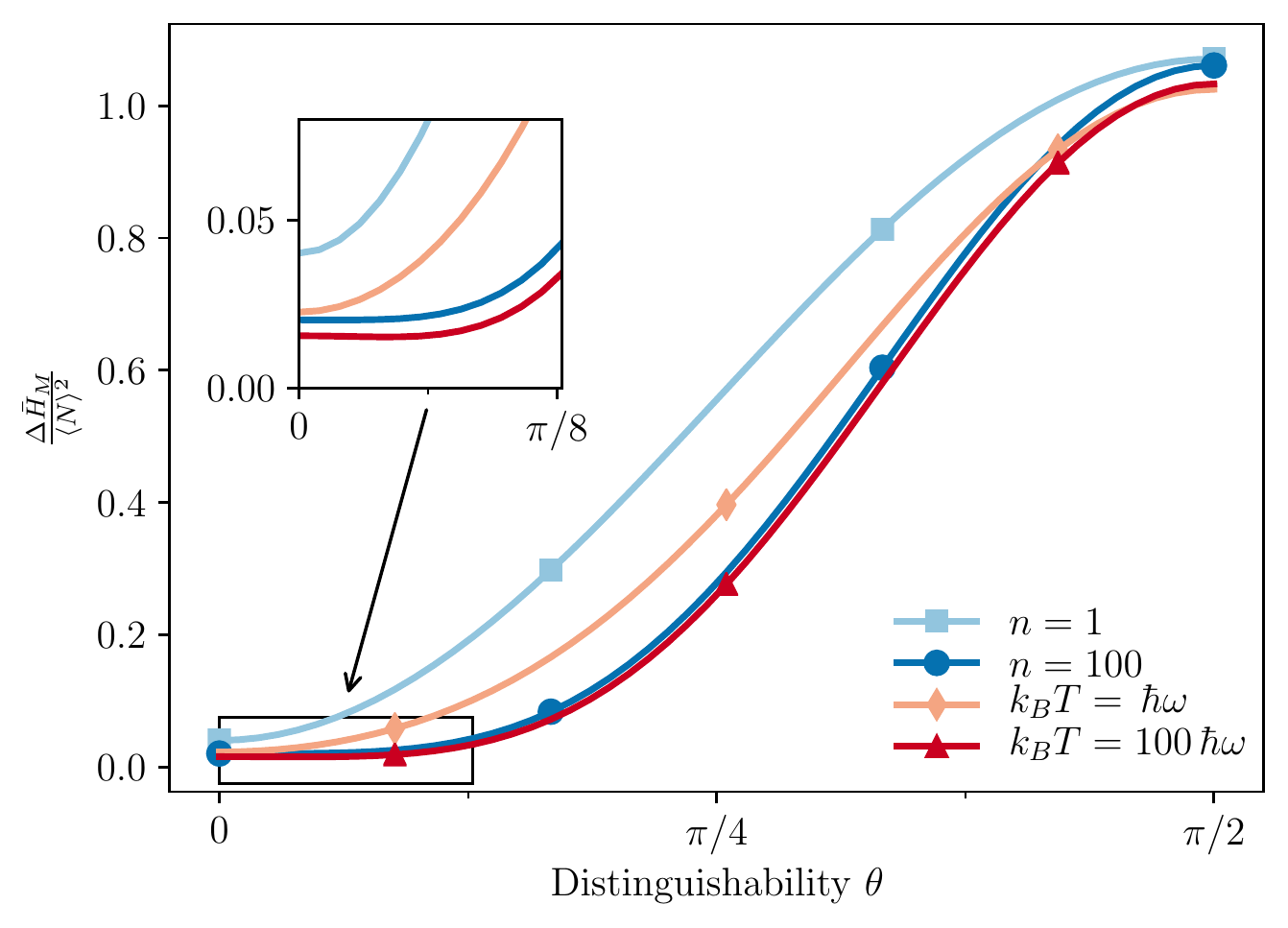}} \hspace{5mm}
\subfloat[Transition from PBS to BS]{\includegraphics[width=0.475\linewidth]{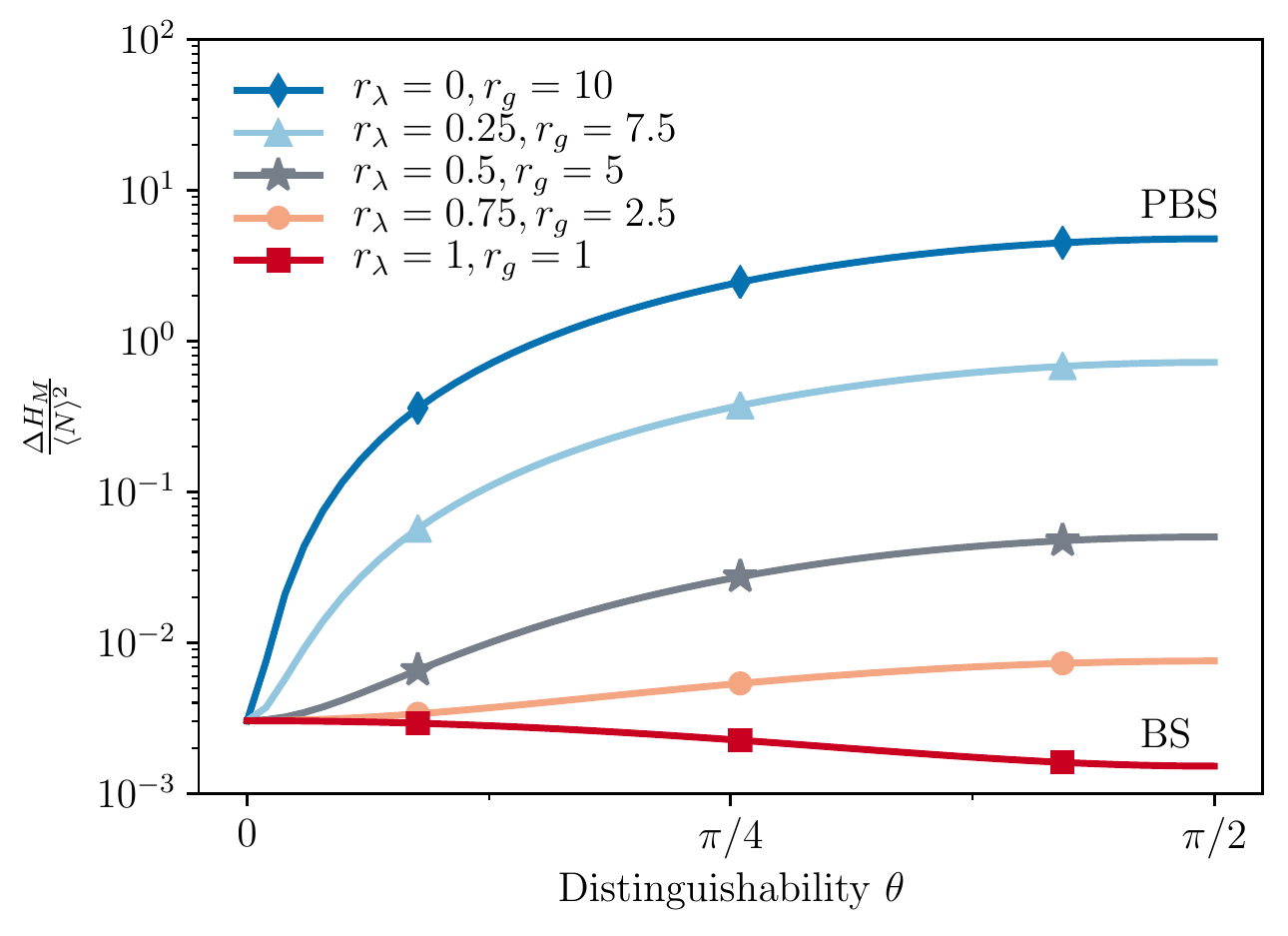}} 
\caption{\label{fig:TotalEnergyPBS}{\bf Energy transfer $\Delta \bar{H}_M$ to PBS membrane.} Time averaged energy transfer to the membrane normalised by the expected number of photons per gas squared, $\Delta \bar{H}_M/\langle N^2 \rangle $, to first order in $g_H$ and $g_V$, as a function of the distinguishability $\theta$ of the photon gases. (a) $\Delta \bar{H}_M/\langle N^2 \rangle $ for a PBS membrane is plotted for photon gases initially in a Fock state with $n=1$ photons per cavity (light blue squares) and $n=100$ photons per cavity (dark blue circles), and for photon gases initially in a thermal state at temperature $k_B T = \hbar \omega $ (pink diamonds) and $k_B T = 100 \hbar \omega$ (red triangles). The inset magnifies the data for $\theta \leq \pi/8$ to highlight that the energy transfer does not vanish as $\theta \to 0$. The high number Fock state (dark blue circles) and high temperature thermal state (red triangles) energy transfers to the membrane converge to the same curve. In this plot $\lambda/\omega_M  = 2$, $g_H/g_V = 6$, $ \omega = 10 \omega_M$ and the energy is given in units of $\frac{\hbar^2 g_H^2}{m \omega_M^2}$. (b) The transition between perfect PBS and BS membranes is indicated by varying $r_\lambda = \frac{\lambda_H}{\lambda_V}$ and $r_g = \frac{g_H}{g_V}$. For $r_g = 1$ and $r_\lambda = 1$ (red squares) the membrane is a perfect BS, whereas for $r_\lambda=0$ and $r_g = 10$ (dark blue diamonds) the membrane is a strongly polarisation dependent beamsplitter. Here $\lambda_V = 4 \omega_M$ and $g_V = 0.1 \omega_M$, the photon gases are initially in a single photon Fock state and the energy is given in units of $\frac{\hbar^2 g_V^2}{m \omega_M^2}$.}
\end{figure}

Since the radiation pressure exerted by the horizontally polarised photons is greater than that exerted by the vertically polarised photons, $g_H > g_V$, the first term of Eq.~\eqref{eq:MembraneFluctTimeAverage} dominates the expression and the time averaged change in energy of the membrane can be approximated as \begin{equation}\label{Eq: W2to1stOrder}
     \Delta \bar{H}_M  \approx  \frac{\hbar^2 g_H^2}{ m \omega_M^2} \langle \Delta N_H^2(0) \rangle \ ,
\end{equation}
where
\begin{equation}\label{eq:DeltaNVsq}
     \langle \Delta N_H^2(0) \rangle = \langle N^2 \rangle \sin^4(\theta)  + \langle N \rangle  \frac{\sin(2\theta)}{4}  \; .
\end{equation}
For both thermal and Fock state photon gases, the value of $\langle N^2 \rangle$ is always greater than $\langle N \rangle$ and therefore the above expression increases monotonically with $\theta$. Thus, as shown in Fig.~\ref{fig:TotalEnergyPBS}(a), the energy transfer to the membrane, similarly to the work output, increases continuously with the distinguishability $\theta$ of the photon gases. Similarly,  as shown in  Fig.~\ref{fig:tempdependence}, and inline with the work output, the energy transfer increases linearly with temperature in the high temperature limit.

The above approximation is valid for large enough $\theta$, but when $\theta \to 0$ then $\langle \Delta N_H^2(0) \rangle$ and $\langle \Delta N_V(0) \Delta N_H(0)  \rangle$ vanish and the $\langle \Delta N^2_V(0)  \rangle$ and $\langle  \Delta K_V^2(0) \rangle$ terms in \eqref{eq:MembraneFluctTimeAverage} become important. These are non-zero implying that the energy transfer to the membrane is strictly non-zero even for the `mixing' of indistinguishable gases. This phenomenon, arises from the fact that the PBS membrane is a microscopic quantum system and thus experiences heating from the radiation pressure exerted by the photons in the cavity irrespective of their polarisation. While, as indicated in Figs.~\ref{fig:tempdependence} and \ref{fig:TotalEnergyPBS}, the contribution of these terms are small compared to the energy transfer for distinguishable gases, such small energy changes may be detected with state-of-the-art experimental techniques~\cite{CooledMembrane}.

\medskip

The energy transfer to a \textit{standard} beamsplitter membrane (BS), as opposed to the \textit{polarising} beamsplitter membrane considered here, was studied in \cite{BunchingOptMech}. There the energy transfer was found to increase with the \textit{indistinguishability} of the photon gases~\cite{BunchingOptMech} due to photonic bunching. Since the PBS considered here also acts as a beamsplitter for $V$-polarised photons, photonic bunching between the $V$ and $\theta$-polarised photon gases also occurs and  contributes to the energy transfer to the membrane for any $\theta \neq \pi/2$. 

The transition between these two regimes can be probed by replacing $H_{\rm PBS}$ with a generalised polarisation dependent beamsplitter interaction Hamiltonian 
\begin{equation}
\begin{aligned}\label{eq:gPBS}
  H_{\mathrm{gPBS}}  = \, & \frac{\hbar \lambda_H}{2} \left(R_H^\dagger L_H + L_H^\dagger R_H \right)  + \frac{\hbar \lambda_V}{2} \left(R_V^\dagger L_V + L_V^\dagger R_V \right)\, .
\end{aligned}
\end{equation} 
A perfect BS is modelled by setting $\lambda_H = \lambda_V$ in Eq.~\eqref{eq:gPBS}, as well as $g_H = g_V$ in $H_{\rm int}$, Eq.~\eqref{eq:OptMechH}, such that the membrane does not differentiate between $H$ and $V$ photons. Conversely for a strongly polarisation dependent beamsplitter $\lambda_H = 0$ and $g_H > g_V$. As shown in red in Fig.~\ref{fig:TotalEnergyPBS}(b), for a perfect BS the energy transfer to the membrane increases with indistinguishability. However, when the ratios of $\lambda_V/\lambda_H$ and $g_H/g_V$ are increased the membrane increasingly acts as a PBS membrane. Then the effect of Gibbs mixing becomes significant, and the magnitude of the energy transfer to the membrane for distinguishable photons increases substantially. Consequently, as shown in dark blue, the effect of bunching is drowned out and the energy transfer to a PBS membrane increases with distinguishability.

\medskip

A deeper understanding of the nature of the energy transfer can be built by splitting it into its contributions from the work output and from fluctuations. Namely, since the time averaged momentum of the membrane vanishes, $\langle P_M \rangle = 0$, the time averaged energy of the membrane can be written as
\begin{equation}
      \langle H_M\rangle  =  W_M + \frac{m \omega^2 \delta X_M}{2} + \frac{\delta P_M }{2 m} \, .
\end{equation}
Here $\frac{m \omega^2 \delta X_M}{2}$, where $\delta X_M$ is the time averaged fluctuations in the displacement membrane, captures the fluctuations in the work output and $\frac{\delta P_M}{2 m} $ captures the heat-like energy associated with the time averaged fluctuations in the momentum of the membrane $\delta P_M$. 
For the case of the BS membrane or indistinguishable H photon gases with a PBS membrane, $W_M$ vanishes and the energy transfer to the membrane is purely fluctuating in origin. However, for distinguishable photons with a PBS membrane, $W_M$ is large and therefore dominates the energy transfer to the membrane.

\section{Experimental Prospects}\label{sec:GibbsExp}

The Fock and thermal configurations were chosen to maintain a close resemblance with the classical Gibbs mixing setting; however, both configurations have experimental limitations. 
Previous `membrane-in-the-middle' experiments operate at cryogenic conditions~\cite{experiment1} which sets the membrane temperature to a few K. Having a thermal photon  state at the same temperature, for a typical cavity frequency $\omega$ of the order of THz, implies low photon numbers, $\langle N\rangle \ll 1$, resulting in an un-measurably small mixing work. 
This could be mitigated by increasing the temperature and hence $\langle N\rangle$  of the thermal photon state; however, this will create an additional  temperature gradient between the photons and the membrane, and so care would need to be taken to ascertain whether or not the work output could be attributed solely to mixing. 
For the Fock state photon gases, \eqref{eq: fockstate}, the challenge is creating a Fock state large enough to have a non-negligible impact on the membrane dynamics or engineering an optomechanical coupling that is sensitive to small numbers of photons. While Fock states of high $n$, e.g., $n  \approx 50$,  remain experimentally demanding, the regime of low photon number optomechanics may be achievable in the near future~\cite{SinglePhoton}.

\begin{figure}[t]
\centering
\includegraphics[width=0.95\linewidth]{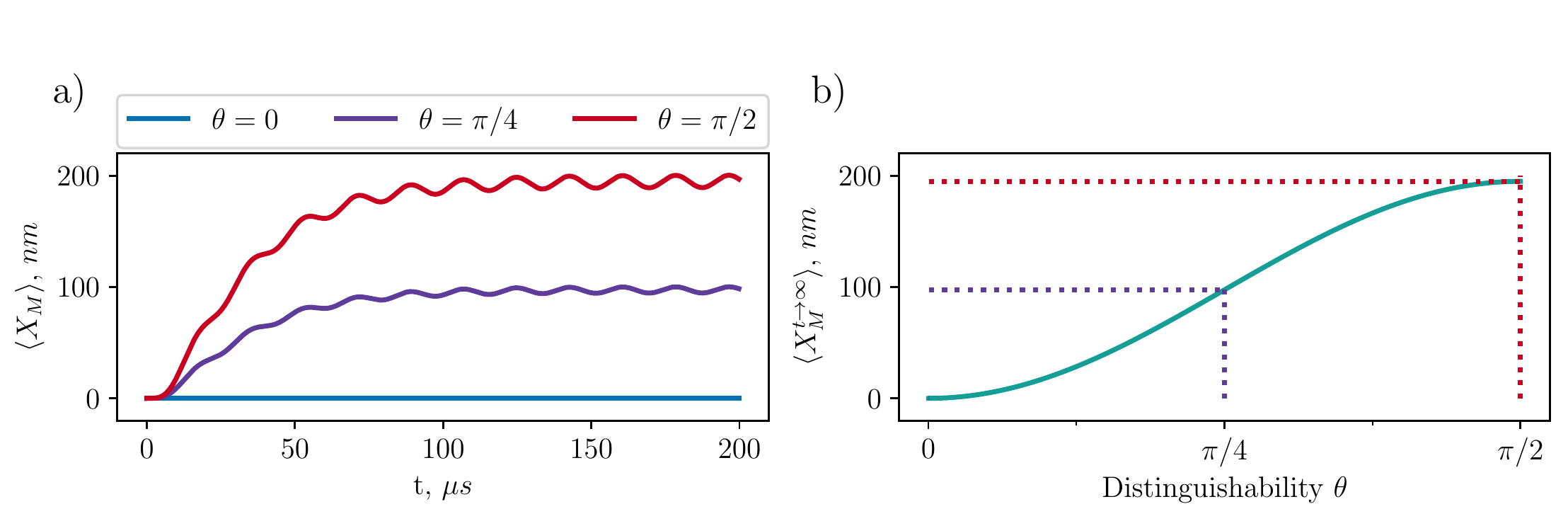}
\caption{\label{fig:PBSimp}{\bf Displacement $ \langle X_M \rangle $, of PBS membrane under proposed implementation.}
a) Expectation value of the displacement of the membrane as a function of time $t$ when the left and right cavities are driven. Polarisation angles $\theta = \pi/2$, $\theta =\pi/4$ and $\theta = 0$ are shown in blue, purple and red respectively. 
b) Estimate for the steady state displacement of the membrane, Eq.~\eqref{eq:DrivenDisEst}, as a function of distinguishability $\theta$.
The steady state displacements in b) for $\theta = 0$, $\theta = \pi/4$ and $\theta =\pi/2$ align well with the long time displacements in a) as indicated by the dotted lines. Since parameters for PBS optomechanical systems are not available  the following experimental parameters from a standard beamsplitter `membrane-in-the-middle' setup~\cite{experiment2, experiment5} have been used: $\omega_M =350$kHz, $\omega =20$THz, $\lambda=34$GHz, $L=93$mm, $\kappa=$85kHz, $\kappa_M=$1Hz, $m=$45ng, $g_V x_{\rm zpf}=$3.3kHz, $g_H x_{\rm zpf}=$19.8kHz and $\epsilon=$40GHz.}
\end{figure}

In this section we will discuss experimental strategies beyond the Fock and thermal configurations whereby the two halves of the cavity are driven by lasers. Any experimental realisation would in practice experience dissipation of photons from the cavity and damping of the mechanical oscillator, and these effects are also incorporated in what follows.

Specifically, we propose simultaneously strongly driving the left cavity with $V$-polarised photons and driving the right cavity with $\theta$-polarised photons. Assuming that this driving is performed on resonance with the cavity frequency and that the two driving processes are in phase,  this can be modelled by adding the following driving term to the Hamiltonian,
\begin{equation}
    H_D = \hbar \epsilon \left((L_V +  R_\theta )\exp( i \omega t) + (L_V^\dagger+R_\theta^\dagger) \exp(- i \omega t)\right) ,
\end{equation}
where $R_\theta = \cos(\theta) R_V + \sin(\theta) R_H$ and $\epsilon$ is the laser driving amplitude~\cite{optomechreview}.
In a frame rotating at the driving frequency, the equation of motion of the photonic modes take the form 
\begin{align}\label{eq:eomPhotons}
    &\frac{d L_V}{dt} = -i \left( (g_V X_M - i \kappa) L_V + \frac{\lambda}{2} R_V + \epsilon \right) \\
    &\frac{d R_p}{dt} = -i \left( -( g_p X_M + i \kappa) R_p + \frac{\lambda_p}{2} L_p + \epsilon_p \right) \; ,
\end{align}
for $p = H, V$ with $\epsilon_V = \epsilon \cos(\theta)$, $\epsilon_H = \epsilon \sin(\theta)$, $\lambda_H = 0$, $\lambda_V =\lambda$ and where $\kappa$ is the rate of cavity dissipation~\cite{optomechreview, GENERALIZEDOPTOMECHANICS,AtomicEnsembMITM}\footnote{Here the damping term is included phenomenologically and we adopt a mean-field approach for the driving term. Strictly the equations of motion include a stochastic noise term of the form $\kappa \, a_{\rm in}(t)$; however, in the limit of large photon and phonon numbers the contribution of noise is expected to be small and so we consider the average input driving term $\epsilon = \langle \kappa a_{\rm in}(t) \rangle$. This suffices to build a coarse-grained  picture of the expected work output in this driven setting.}. Similarly, the equation of motion of the membrane, cf. Eq.\eqref{Eq:MotionMembrane}, now reads
\begin{align}\label{eq:eomMembrane}
    \frac{d^2 X_M}{d t^2} +   \kappa_M \frac{d X_M}{d t} + \omega_M^2 X_M  =  \frac{F(g_H, g_V)}{m} \; ,   
\end{align}
where we account for the damping of the mechanical mode at rate $\kappa_M$~\cite{optomechreview, GENERALIZEDOPTOMECHANICS,AtomicEnsembMITM}. Fig.~\ref{fig:PBSimp} shows the average displacement of the membrane as a function of time, $ \langle X_M \rangle $,  found by solving these differential equations to first order in $g_H$ and $g_V$ and assuming that the left and right halves of the cavity are initially in the vacuum state. The displacement of the membrane is largest for the mixing of perfectly distinguishable photon gases ($\theta = \pi/2$, red), vanishes for perfectly indistinguishable gases ($\theta = 0$, blue), and sitting between these two extremes is the displacement for partially distinguishable gases (e.g. $\theta = \pi/4$, purple).

In the limit\footnote{This limit has been realised experimentally for standard beamsplitter membranes~\cite{experiment2}. As \textit{polarisation dependent} beamsplitter membranes are yet to be utilised in an optomechanical experimental setup, their damping rates are not currently known.} 
in which the cavity damping rate $\kappa$ is substantially faster than the membrane damping rate $\kappa_M$, the displacement of the membrane increases initially before reaching a steady oscillatory state about a new displaced origin on the time scale $\frac{1}{\kappa} \ll t \ll \frac{1}{\kappa_M}$. This new displaced origin is found by disregarding the effect of membrane damping altogether, and evaluating the displacement, averaged over fast oscillations, in the limit of large $t$. The steady state displacement of the membrane is found to obey
\begin{equation}\label{eq:DrivenDisEst}
     \lim_{t \gg \frac{1}{\kappa}} \langle X_M(t) \rangle \approx  \frac{4 \epsilon^2 \hbar g_H}{m \kappa^2 \omega_M^2} \sin^2(\theta) \, ,
\end{equation}
in the limit that $g_H \gg g_V$. 
For the parameters used \cite{experiment2} the full numerical displacements given by \eqref{eq:TimeAverageDisplacement}, and shown in Fig.~\ref{fig:PBSimp}a) for $\theta = 0$, $\theta = \pi/4$ and $\theta =\pi/2$, align well at long times with the predicted steady state displacements given by \eqref{eq:DrivenDisEst} and shown in Fig.~\ref{fig:PBSimp}b).
Thus the displacement for this driven damped implementation is directly proportional to the displacement discussed in Section~\ref{sec:OptMechMixResults} for the un-damped and un-driven case, Eq.~\eqref{eq:TimeAverageDisplacement}, increasing continuously with the distinguishability of the photon gases.  In contrast to the demanding Fock and thermal states, this driven implementation provides a more viable experimental scheme to observe a continuous increase in the mixing work as a function of the distinguishability of two polarised photon gases.

While `membrane-in-the-middle' experiments using beamsplitter membranes are well established~\cite{experiment1,experiment1a,experiment2,experiment3,experiment4,experiment5}, similar setups utilising polarisation dependent beamsplitters have not yet been investigated. As a result, experimental values for $g_H$, $g_V$, $m$, $\omega_M$ and $\kappa_M$ for ultra-thin PBS membranes are not readily available and this lack of data makes it hard to fully access the viability of our proposal. 
The smallest PBS membranes currently engineered are a couple of orders of magnitude thicker than the membranes that are typically used in `membrane-in-the-middle' experiments~\cite{experiment5, ThinPBS1, ThinPBS2}, suggesting that it might be challenging in the short term to realise quantum regimes of the `PBS-in-the-middle' setting. However, this is an unknown since it is yet to be tried. Moreover, active progress in miniaturising PBS membranes~\cite{ThinPBS1, ThinPBS2} will continue to increase the feasibility of our proposal.

\section{Conclusion and Outlook} \label{sec:conc}

In this paper we have proposed an optomechanical setup with a  `PBS-in-the-middle' of a cavity separating two homogeneous photon gases, one $V$-polarised and the other $\theta$-polarised. By varying $\theta$ we can smoothly vary the distinguishability of the two gases allowing us to explore not only the classical limits of two gases that are perfectly distinguishable or perfectly indistinguishable but also consider quantum gases, that in contrast to the classical case, are fundamentally only partially distinguishable.
The setup provides a natural physical system where the work from mixing two gases is stored: the potential energy associated with the displacement to the microscopic membrane. 
Using the Hamiltonian equations of motion for this physically motivated setup we have derived a formula for the work drawn from mixing the two gases as a function of their polarisation-distinguishability $\theta$, and valid for any number distribution of both photon gases. 

The mixing work here varies smoothly with the polarisation distinguishability inline with previous results \cite{thesoviets,allahverdyan}. \zo{While there is a classical model, namely one using inhomogeneous gases composed of a mixture of polarisation states, that is capable of producing the smooth change in the output with distinguishability, this deviates from the spirit of the original Gibbs mixing thought experiment which applies to the mixing of homogeneous gases. Crucially, there is no classical model involving only homogeneous gases that can explain the smooth output with distinguishability observed in the optomechanical setting considered here.}

The optomechanical setting additionally generalises Gibbs mixing to the quantum regime in the sense that the PBS interaction gives the photons access to coherent superpositions between the two halves of the cavity, and the optomechanical interaction entangles the photonic modes and the microscopic piston. 
% Given that the $\theta$ photons are in a superposition of V and H states, and the PBS is only permeable to H photons, the coherent interaction of the gases with the PBS necessarily generates spatial coherence. In this sense, the coherence of the spatial degree of freedom is essential to the mixing process.  
\zo{Since a $V$ photon impinging on the PBS can be transmitted while a $H$ photon is always reflected, the coherent action of the PBS on a photon with a $\theta$ polarisation (superposition of $V$ and $H$) will be coherently transmitted as well as reflected by the PBS. In this manner, the PBS converts coherence in the polarisation degree of freedom into coherence in the spatial degree of freedom. Given that the possibility of smoothly varying the polarisation is key to be able to consider quantum Gibbs mixing, the spatial coherence goes hand in hand with it in this optomechanical setting.}
Moreover, the entanglement between the photons and the membrane is crucial to understanding the behaviour of the total energy transfer to the membrane and the work fluctuations. Specifically, due to the back action of photon bunching on the membrane, a fluctuating component of the energy transfer to the membrane increases with indistinguishability.
Thus, intriguingly, the effects of bunching and Gibbs mixing are antithetical. Nonetheless, since the energy transfer to the membrane due to bunching is small in comparison to that due to quantum Gibbs mixing, the overall energy transfer to the PBS is expected to increase as a function of distinguishability.

% The fluctuations of the mixing work are not negligible for the microscopic piston and, as we illustrate for Fock and thermal photon number distributions, they are largest at low photon numbers/low temperatures where vacuum fluctuations dominate thermal fluctuations.

In a driven and dissipative cavity setting, the membrane displacement plateaus to a constant steady state that increases smoothly with $\theta$.
While previous entropic and information-theoretic derivations of the mixing work~\cite{Gibbs,thesoviets,allahverdyan} have provided thermodynamic bounds on the maximal work extraction, the driven optomechanical cavity setting discussed here could conceivably be realised experimentally in the near future offering the chance to measure the continuous quantum mixing work for the first time,  without needing to perform an optimal protocol. Nonetheless, it would be interesting to investigate if it is possible to come up with variations of physically realisable schemes that may extract more work from mixing the same gases. One could, for example, consider variants of the setup utilising multiple PBS membranes to enable the two photon gases to mix fully rather than partially as currently. Alternatively, one could consider using initial gases that are entangled with respect to the polarisation degree of freedom, for example W or GHZ states, to mimic the quantum enhancements found in other applications such as parameter estimation~\cite{GHZparameterestimation} and measurement based work extraction protocols~\cite{GHZworkextract}.

The `PBS-in-the-middle' optomechanical setup also provides a framework to further study the mixing of photon gases utilising alternative types of membranes \cite{BunchingOptMech}.
In addition to the semi-permeable membrane used in Gibbs mixing, many of the pioneering thought experiments in thermodynamics can be framed in terms of gases performing work on a membrane attached to a movable piston. For example, Feynman's ratchet~\cite{FeynmanRatchetOrig} (a variant of the Szilard-experiment~\cite{Szilard}) effectively uses a uni-directionally transmissive membrane and a Maxwell demon~\cite{Maxwell} could take the form of a fictitious semi-permeable membrane that separates fast and slow moving particles. This suggests that `membrane-in-the-middle' optomechanics has wide ranging potential, not just as a platform to study the role of distinguishability and mixing in thermodynamics, but also for probing the fundamental relationships between information, heat and work. 

We note that in the present optomechanical setup temperature enters through the initial thermal state of the membrane, and optionally as the temperature of initially thermal photon gases, but no continual thermalisation with a heat bath is modelled during the mixing process. The continual thermalisation of Gibbs' classical gases could be studied in the current setting either through a more detailed analysis of the dissipative coupling of the cavity modes to their surrounding thermal environment, or by using dye-molecules to actively mediate effective interactions between photons resulting in thermalisation~\cite{klaers2010,tinyBEC}. Since the mechanical and light modes interact with different reservoirs, temperature gradients between the gases and the piston could be introduced that we expect will lead to additional quantum thermodynamic effects.

\bigskip

\textit{Note: Since completing of our paper, another work has appeared on the arXiv~\cite{GibbsMixBens}, which provides a new information theoretic analysis of Gibbs mixing in the quantum regime that takes into account the knowledge of an observer, while not detailing a system’s Hamiltonian or time-evolution. }

\bigskip
\bigskip

{\bf Acknowledgements.} 
We are grateful for insightful conversations with Jack Clarke, Michael Vanner and Alexia Auffeves.
We acknowledge support from the Engineering and Physical Sciences Research Council Centre for Doctoral Training in Controlled Quantum Dynamics; the Engineering and Physical Sciences Research Council Grants EP/M009165/1, EP/R045577/1  and EP/S000755/1, and the Royal Society.

\bigskip

\bibliographystyle{ieeetr} 
\bibliography{refs}

\begin{thebibliography}{10}

\bibitem{Gibbs}
J.~Gibbs, ``On the equilibrium of heterogeneous substances,'' {\em Connecticut
  Acad. Sci.}, vol.~3, p.~343–524, 1875-1878.

\bibitem{Schrodinger}
E.~Schr\"odinger, {\em Statistical Thermodynamics}.
\newblock Cambridge University Press, 1952.

\bibitem{thesoviets}
V.~Luboshitz and M.~Podgoretskii, ``The gibbs paradox,'' {\em Sov. Phys. Usp.},
  vol.~14, p.~662, 1972.

\bibitem{Lande1}
A.~Land\`e, {\em New Foundations of Quantum Mechanics}.
\newblock Cambridge University Press, 1926.

\bibitem{Lande2}
A.~Land\`e, {\em Foundations of Quantum Theory}.
\newblock Yale University Press, 1952.

\bibitem{allahverdyan}
A.~E. Allahverdyan and T.~M. Nieuwenhuizen, ``Explanation of the gibbs paradox
  within the framework of quantum thermodynamics,'' {\em Phys. Rev. E},
  vol.~73, p.~066119, Jun 2006.

\bibitem{Peres}
A.~Peres, {\em Quantum Theory: Concepts and Methods}.
\newblock Springer, 1995.

\bibitem{Maruyama}
K.~Maruyama, {\v{C}}.~Brukner, and V.~Vedral, ``Thermodynamical cost of
  accessing quantum information,'' {\em Journal of Physics A: Mathematical and
  General}, vol.~38, pp.~7175--7181, Jul 2005.

\bibitem{ergotropy}
A.~E. Allahverdyan, R.~Balian, and T.~M. Nieuwenhuizen, ``Maximal work
  extraction from finite quantum systems,'' {\em Europhysics Letters ({EPL})},
  vol.~67, pp.~565--571, Aug 2004.

\bibitem{Elouard_2015}
C.~Elouard, M.~Richard, and A.~Auff{\`{e}}ves, ``Reversible work extraction in
  a hybrid opto-mechanical system,'' {\em New Journal of Physics}, vol.~17,
  p.~055018, may 2015.

\bibitem{OptoMechExperiment}
M.~Brunelli, L.~Fusco, R.~Landig, W.~Wieczorek, J.~Hoelscher-Obermaier,
  G.~Landi, F.~L. Semi\~ao, A.~Ferraro, N.~Kiesel, T.~Donner, G.~De~Chiara, and
  M.~Paternostro, ``Experimental determination of irreversible entropy
  production in out-of-equilibrium mesoscopic quantum systems,'' {\em Phys.
  Rev. Lett.}, vol.~121, p.~160604, Oct 2018.

\bibitem{OptMechthermo1}
H.~Ian, ``Thermodynamic cycle in a cavity optomechanical system,'' {\em Journal
  of Physics B: Atomic, Molecular and Optical Physics}, vol.~47, p.~135502, jun
  2014.

\bibitem{OptoMechthermo2}
C.~Elouard, M.~Richard, and A.~Auff{\`{e}}ves, ``Reversible work extraction in
  a hybrid opto-mechanical system,'' {\em New Journal of Physics}, vol.~17,
  p.~055018, may 2015.

\bibitem{OptoMechthermo3}
J.~Monsel, C.~Elouard, and A.~Auff{\`e}ves, ``An autonomous quantum machine to
  measure the thermodynamic arrow of time,'' {\em npj Quantum Information},
  vol.~4, no.~1, p.~59, 2018.

\bibitem{OptoMechthermo4}
M.~{Konopik}, A.~{Friedenberger}, N.~{Kiesel}, and E.~{Lutz}, ``{Nonequilibrium
  information erasure below kTln2},'' {\em arXiv e-prints},
  p.~arXiv:1806.01034, Jun 2018.

\bibitem{OptoMechthermo5}
J.~S. Bennett, L.~S. Madsen, H.~Rubinsztein-Dunlop, and W.~P. Bowen, ``A
  quantum heat machine from fast optomechanics,'' {\em New Journal of Physics},
  vol.~22, p.~103028, oct 2020.

\bibitem{experiment1}
J.~D. Thompson, B.~M. Zwickl, A.~M. Jayich, F.~Marquardt, S.~M. Girvin, and
  J.~G.~E. Harris, ``Strong dispersive coupling of a high-finesse cavity to a
  micromechanical membrane,'' {\em Nature}, vol.~452, p.~72, 03 2008.

\bibitem{experiment1a}
A.~M. Jayich, J.~C. Sankey, B.~M. Zwickl, C.~Yang, J.~D. Thompson, S.~M.
  Girvin, A.~A. Clerk, F.~Marquardt, and J.~G.~E. Harris, ``Dispersive
  optomechanics: a membrane inside a cavity,'' {\em New Journal of Physics},
  vol.~10, p.~095008, Sep 2008.

\bibitem{experiment2}
M.~Karuza, C.~Biancofiore, M.~Bawaj, C.~Molinelli, M.~Galassi, R.~Natali,
  P.~Tombesi, G.~Di~Giuseppe, and D.~Vitali, ``Optomechanically induced
  transparency in a membrane-in-the-middle setup at room temperature,'' {\em
  Phys. Rev. A}, vol.~88, p.~013804, Jul 2013.

\bibitem{experiment3}
J.~C. Sankey, C.~Yang, B.~M. Zwickl, A.~M. Jayich, and J.~G.~E. Harris,
  ``Strong and tunable nonlinear optomechanical coupling in a low-loss
  system,'' {\em Nature Physics}, vol.~6, pp.~707 EP --, 06 2010.

\bibitem{experiment4}
D.~Lee, M.~Underwood, D.~Mason, A.~B. Shkarin, S.~W. Hoch, and J.~G.~E. Harris,
  ``Multimode optomechanical dynamics in a cavity with avoided crossings,''
  {\em Nature Communications}, vol.~6, pp.~6232 EP --, 02 2015.

\bibitem{experiment5}
M.~Karuza, M.~Galassi, C.~Biancofiore, C.~Molinelli, R.~Natali, P.~Tombesi,
  G.~D. Giuseppe, and D.~Vitali, ``Tunable linear and quadratic optomechanical
  coupling for a tilted membrane within an optical cavity: theory and
  experiment,'' {\em Journal of Optics}, vol.~15, p.~025704, Dec 2012.

\bibitem{experiment6}
C.~Reinhardt, T.~M\"uller, A.~Bourassa, and J.~C. Sankey, ``Ultralow-noise sin
  trampoline resonators for sensing and optomechanics,'' {\em Phys. Rev. X},
  vol.~6, p.~021001, Apr 2016.

\bibitem{BunchingOptMech}
Z.~Holmes, J.~Anders, and F.~Mintert, ``Enhanced energy transfer to an
  optomechanical piston from indistinguishable photons,'' {\em Phys. Rev.
  Lett.}, vol.~124, p.~210601, May 2020.

\bibitem{JaynesGibbs}
E.~T. Jaynes, ``The gibbs paradox,'' in {\em Maximum Entropy and Bayesian
  Methods} (J.~Skilling, ed.), Kluwer Acad. Pub., 1991.

\bibitem{SaundersGibbs}
S.~Saunders, ``The gibbs paradox,'' {\em Entropy}, vol.~20, p.~552, 2018.

\bibitem{DieksGibbs}
D.~Dieks, ``The gibbs paradox and particle individuality,'' {\em Entropy},
  vol.~20, p.~466, 2018.

\bibitem{zeilinger}
G.~Weihs and A.~Zeilinger, ``{Photon statistics at beam-splitters: an essential
  tool in quantum information and teleportation},'' in {\em Coherence and
  Statistics of Photons and Atoms} (J.~Perina, ed.), Wiley and Wiley,, 2001.

\bibitem{polarisationOptomech}
H.~Xiong, Y.-M. Huang, L.-L. Wan, and Y.~Wu, ``Vector cavity optomechanics in
  the parameter configuration of optomechanically induced transparency,'' {\em
  Phys. Rev. A}, vol.~94, p.~013816, Jul 2016.

\bibitem{RadiationPressure}
S.~Chandrasekhar, {\em Radiative Transfer}.
\newblock Oxford University Press, 1950.

\bibitem{defofquantumwork1}
A.~E. Allahverdyan and T.~M. Nieuwenhuizen, ``Fluctuations of work from quantum
  subensembles: The case against quantum work-fluctuation theorems,'' {\em
  Phys. Rev. E}, vol.~71, p.~066102, Jun 2005.

\bibitem{defofquantumwork2}
P.~Talkner, E.~Lutz, and P.~H\"anggi, ``Fluctuation theorems: Work is not an
  observable,'' {\em Phys. Rev. E}, vol.~75, p.~050102, May 2007.

\bibitem{defofquantumwork3}
P.~Talkner and P.~H\"anggi, ``Aspects of quantum work,'' {\em Phys. Rev. E},
  vol.~93, p.~022131, Feb 2016.

\bibitem{defofquantumwork3.5}
P.~Kammerlander and J.~Anders, ``Coherence and measurement in quantum
  thermodynamics,'' {\em Scientific Reports}, vol.~6, p.~22174, Feb 2016.

\bibitem{defofquantumwork4}
M.~Perarnau-Llobet, E.~B\"aumer, K.~V. Hovhannisyan, M.~Huber, and A.~Acin,
  ``No-go theorem for the characterization of work fluctuations in coherent
  quantum systems,'' {\em Phys. Rev. Lett.}, vol.~118, p.~070601, Feb 2017.

\bibitem{defofquantumwork5}
M.~Lostaglio, ``Quantum fluctuation theorems, contextuality, and work
  quasiprobabilities,'' {\em Phys. Rev. Lett.}, vol.~120, p.~040602, Jan 2018.

\bibitem{CooledMembrane}
T.~P. Purdy, R.~W. Peterson, P.-L. Yu, and C.~A. Regal, ``Cavity optomechanics
  with si3n4membranes at cryogenic temperatures,'' {\em New Journal of
  Physics}, vol.~14, p.~115021, Nov 2012.

\bibitem{SinglePhoton}
A.~Nunnenkamp, K.~B\o{}rkje, and S.~M. Girvin, ``Single-photon optomechanics,''
  {\em Phys. Rev. Lett.}, vol.~107, p.~063602, Aug 2011.

\bibitem{optomechreview}
M.~Aspelmeyer, T.~J. Kippenberg, and F.~Marquardt, ``Cavity optomechanics,''
  {\em Rev. Mod. Phys.}, vol.~86, pp.~1391--1452, Dec 2014.

\bibitem{GENERALIZEDOPTOMECHANICS}
K.~D.~Z. Li~Jin-Jin, {\em Generalized Optomechanics and its applications:
  quantum optical properties of generalised optomechanical system}.
\newblock World Scientific, 2013.

\bibitem{AtomicEnsembMITM}
A.~K. Chauhan and A.~Biswas, ``Motion-induced enhancement of rabi coupling
  between atomic ensembles in cavity optomechanics,'' {\em Phys. Rev. A},
  vol.~95, p.~023813, Feb 2017.

\bibitem{ThinPBS1}
R.~M.~A. Azzam, ``Simplified design of thin-film polarizing beam splitter using
  embedded symmetric trilayer stack,'' {\em Appl. Opt.}, vol.~50,
  pp.~3316--3320, Jul 2011.

\bibitem{ThinPBS2}
L.~Li and J.~A. Dobrowolski, ``High-performance thin-film polarizing beam
  splitter operating at angles greater than the critical angle,'' {\em Appl.
  Opt.}, vol.~39, pp.~2754--2771, Jun 2000.

\bibitem{GHZparameterestimation}
H.~Yuan and C.-H.~F. Fung, ``Quantum parameter estimation with general
  dynamics,'' {\em npj Quantum Information}, vol.~3, no.~1, p.~14, 2017.

\bibitem{GHZworkextract}
M.~A. Ciampini, L.~Mancino, A.~Orieux, C.~Vigliar, P.~Mataloni, M.~Paternostro,
  and M.~Barbieri, ``Experimental extractable work-based multipartite
  separability criteria,'' {\em npj Quantum Information}, vol.~3, no.~1, p.~10,
  2017.

\bibitem{FeynmanRatchetOrig}
R.~P. Feynman, R.~B. Leighton, M.~Sands, and E.~M. Hafner, {\em The Feynman
  Lectures on Physics; Vol.I}.
\newblock Addison-Wesley, 1965.

\bibitem{Szilard}
L.~Szilard, ``Uber die entropieverminderung in einem thermodynamischen system
  bei eingriffen intelligenter wesen.,'' {\em Z. Phys.}, vol.~53, p.~840–856,
  1929.

\bibitem{Maxwell}
J.~C. Maxwell, {\em Theory of heat}.
\newblock Longmans, Green, and Co., 1908.

\bibitem{klaers2010}
J.~Klaers, J.~Schmitt, F.~Vewinger, and M.~Weitz, ``Bose--einstein condensation
  of photons in an optical microcavity,'' {\em Nature}, vol.~468, pp.~545 EP
  --, 11 2010.

\bibitem{tinyBEC}
B.~T. Walker, L.~C. Flatten, H.~J. Hesten, F.~Mintert, D.~Hunger, A.~A.~P.
  Trichet, J.~M. Smith, and R.~A. Nyman, ``Driven-dissipative non-equilibrium
  bose--einstein condensation of less than ten photons,'' {\em Nature Physics},
  vol.~14, no.~12, pp.~1173--1177, 2018.

\bibitem{GibbsMixBens}
B.~{Yadin}, B.~{Morris}, and G.~{Adesso}, ``{Extracting work from mixing
  indistinguishable systems: A quantum Gibbs ``paradox''},'' {\em arXiv
  e-prints}, p.~arXiv:2006.12482, June 2020.

\end{thebibliography}

\end{document}